\documentclass[aip,graphicx,amsmath,amssymb,reprint,twocolumn]{revtex4-1}
\usepackage[english]{babel}
\usepackage{ucs}
\usepackage[utf8x]{inputenc}
\usepackage{graphicx}
\usepackage{dcolumn}
\usepackage{bm}
\usepackage{pifont}
\usepackage{bbm}
\usepackage{amsmath}
\usepackage{amssymb}
\usepackage{mathrsfs}
\usepackage{bm}
\usepackage{bbold}
\usepackage{empheq}
\usepackage{amsfonts}
\usepackage{verbatim}
\usepackage{color,colordvi}
\usepackage{physics}
\usepackage[colorlinks=true]{hyperref}
\usepackage[capitalize]{cleveref}

\newcommand{\RomanNumeralCaps}[1]{\MakeUppercase{\romannumeral #1}}
\draft % marks overfull lines with a black rule on the right

\begin{document}

% Use the \preprint command to place your local institutional report number 
% on the title page in preprint mode.
% Multiple \preprint commands are allowed.
%\preprint{}

\title{Perfect pulsed inline twin-beam squeezers}

% repeat the \author .. \affiliation  etc. as needed
% \email, \thanks, \homepage, \altaffiliation all apply to the current author.
% Explanatory text should go in the []'s, 
% actual e-mail address or url should go in the {}'s for \email and \homepage.
% Please use the appropriate macro for the type of information

% \affiliation command applies to all authors since the last \affiliation command. 
% The \affiliation command should follow the other information.

\author{Martin Houde}
 \email[]{martin.houde@polymtl.ca.}
% \affiliation{Department of Engineering Physics, École Polytechnique de Montréal, 2500 Chem. de Polytechnique, Montréal, Quebec H3T 1J4, Canada}%Lines break automatically or can be forced with \\
\author{Nicolás~Quesada}%
 \email[]{nicolas.quesada@polymtl.ca.}
\affiliation{ 
D\'epartement de g\'enie physique, \'Ecole Polytechnique de Montr\'eal, Montr\'eal, QC, H3T 1J4, Canada}%

% Collaboration name, if desired (requires use of superscriptaddress option in \documentclass). 
% \noaffiliation is required (may also be used with the \author command).
%\collaboration{}
%\noaffiliation

\date{\today}

\begin{abstract}
    Perfect inline squeezers are both spectrally pure and have identical input and output temporal modes, allowing one to squeeze an arbitrary input quantum state in the sole input mode on which the device acts, while the quantum states of any other modes are unaffected. We study theoretically how to obtain a perfect pulsed inline squeezer in twin-beam systems by considering three commonly used configurations: unpoled single pass, poled single pass, and poled double pass. By obtaining analytical relations between the input and output temporal modes from the Bloch-Messiah decomposition of the discretized Heisenberg-picture propagator, we find that a double pass structure produces a perfect pulsed inline squeezer when operated in a frequency degenerate, symmetric group-velocity matched type-\RomanNumeralCaps{2} configuration.   
\end{abstract}

\pacs{}% insert suggested PACS numbers in braces on next line

\maketitle %\maketitle must follow title, authors, abstract and \pacs
    %%%%%%%%%%%%%%%%%%%%%%%%%%%%%%%%%%%%%%%%%%%%%%%%%%%%%%%%%%%%%%%%%%%%%%%%
	%%%%%%%%%%%%%%%%%%%%%%%%%%%%%%%%%%%%%%%%%%%%%%%%%%%%%%%%%%%%%%%%%%%%%%%%
	\section{Introduction}
	\label{sec:Intro}
	%%%%%%%%%%%%%%%%%%%%%%%%%%%%%%%%%%%%%%%%%%%%%%%%%%%%%%%%%%%%%%%%%%%%%%%%
	%%%%%%%%%%%%%%%%%%%%%%%%%%%%%%%%%%%%%%%%%%%%%%%%%%%%%%%%%%%%%%%%%%%%%%%%

    Squeezed states of light are ubiquitous in quantum information and quantum computing in the optical domain~\cite{braunstein2005info,Weedbrook2012info,braunstein2005squeezing}. 
    Recently, there have been great advancements made on how to generate and use these states as resources in quantum computing both theoretically~\cite{christ2013theory,christ2014theory,Helt2020Degen,quesada2022BPP,quesada2020theory,vernon2019scalable, arzani2018versatile,houde2023sources,grier2022complexity} and experimentally~\cite{zhong2020exp1,zhong2021exp2,zhong2019experimental,eckstein2011highly,wang2019gbs,Xanadu2022gbs,Arrazola2021circuits,triginer2020cascaded,harder2013optimized,harder2016single, Yanagimoto2023nondemo, Rajveer2023Fewcycle,thekkadath2024gaininduced}.

    Much of the research has been geared towards finding squeezers that generate squeezed vacuum states that are bright, single- and consistent- temporal mode, and amenable to be generated in integrated platforms\cite{vernon2019scalable}. However, there is still much work to be done to characterize how these squeezers act as inline squeezers, i.e., when they are seeded by an arbitrary quantum state that might differ from the
    vacuum. Inline squeezers have been shown to be important in non-Gaussian state generation using heralding\cite{crescimanna2023seeding,Furasawa2014dynamicsqueezing,winnel2023deterministic}, quantum metrology using su(1,1) interferometers~\cite{yurke19862,roeder2023measurement,roeder2023measurement}, simulation of quantum field theories~\cite{briceno2023toward} and as continuous-variable quantum computing gates~\cite{braunstein2005info,kalajdzievski2021exact,Weedbrook2012info}
    
	Typical experimental inline squeezing scheme might involve the generation of a twin-beam squeezed vacuum state, heralding the idler such that we are in a known signal state, and sending the heralded signal state back in the squeezer to hopefully generate a squeezed state of light in the heralded state. However, as we outline in Sec.~\ref{sec:Issues}, one must be careful when doing so. Squeezers are characterized by two different sets of Schmidt or temporal modes, the input modes and the output modes. In the case of squeezing vacuum, only the output modes are relevant\cite{quesada2022BPP}, as the vacuum is matched to any input mode. However, in the case of inline squeezing, i.e., seeding the squeezer with a non-vacuum state, both the input and output states are important. If one wants to generate a squeezed state in a specific output mode, one needs to send in the corresponding input mode. To fully characterize a squeezer it is therefore important to study both input and output Schmidt modes. 
    
\begin{figure*}[t]
		\includegraphics[width=\linewidth]{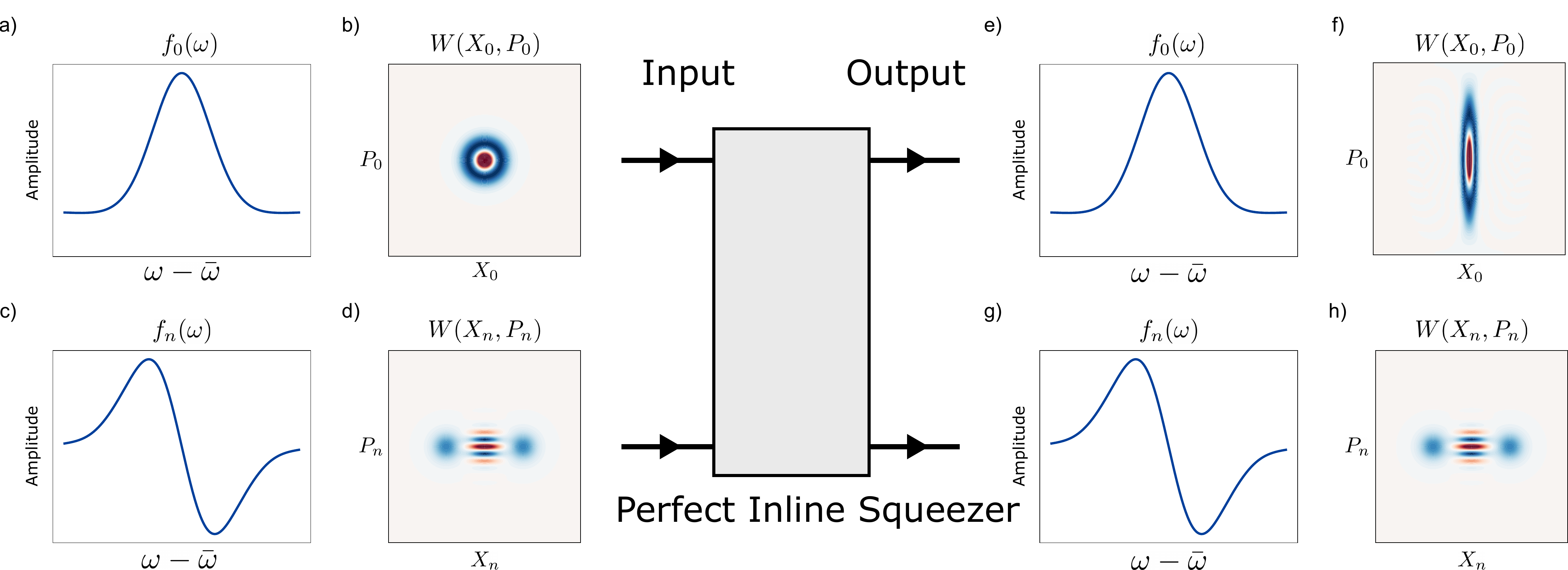}
		\caption{Schematic representation of a perfect inline squeezer. (a) Temporal mode structure with respect to central frequency $\bar{\omega}$, $f_{0}(\omega)$, that is matched to the perfect inline squeezer. (b) Wigner distribution of the injected quantum state, taken to be a single Fock state, for the matched mode. (c) Temporal mode structure with respect to central frequency $\bar{\omega}$, $f_{n}(\omega)$, of an unmatched mode. (d) Wigner distribution of the injected quantum state, taken to be a Cat state, for the unmatched mode. (e) Temporal mode structure of the output state for the matched input state. Note that the structure remains the same. (f) Wigner distribution of the output state for the matched case. The output is now is a Squeezed Fock state. (g) Temporal mode structure of the output state for the unmatched input state. Note that the structure remains the same. (h) Wigner distribution of the output state for the unmatched case. The Wigner distribution remains unchanged. Only the matched input generates a squeezed output.}
		\label{fig:perfect}
	\end{figure*}

    We say that an inline squeezer is perfect when it is both spectrally pure (i.e. has only one output Schmidt mode) and the input and output modes are matched (i.e. the modes are identical). Figure~\ref{fig:perfect} shows a schematic representation of how a perfect inline squeezer functions. For demonstrative purposes, we take Fig.~\ref{fig:perfect}(a) to be the single temporal mode that is mode-matched to the inline squeezer. When light in this mode is injected in the perfect inline squeezer in an arbitrary state (taken to be a single Fock state with Wigner distribution shown in Fig.~\ref{fig:perfect}(b)), the output state generated has the same temporal structure but now has a squeezed Wigner function (see Fig.~\ref{fig:perfect}(f)). When light in any other orthogonal temporal mode (Fig.~\ref{fig:perfect}(c)) in an arbitrary state (Fig.~\ref{fig:perfect}(d)) is injected in the perfect inline squeezer, the state is unaffected by the inline squeezer and as such the output is the same as the input. An imperfect squeezer would generate an output that is either in a different temporal mode (when the input and output modes differ) or in an unwanted mode (when it is not spectrally pure). 
    
    One can gain information on both sets of modes by applying a Bloch-Messiah decomposition to the Heisenberg-picture propagator describing the evolution of the squeezer~\cite{Cariolar2016BM1,Cariolaro2016BM2,mccutcheon2018structure,horoshko2019bloch}. This type of decomposition is readily implemented numerically~\cite{houde2024decomp}, however, it is generally very hard to obtain analytically, especially when a large number of modes are used to approximate the quasi-continuum encountered in waveguide geometries. In this paper, we look at different waveguided twin-beam squeezers, which are experimentally relevant~\cite{zhong2020exp1,zhong2021exp2,zhong2019experimental}, and develop a method for obtaining analytical relations between the input and output modes, without computing them explicitly. 
     By working with discretized equations of motion and using the symmetries of the matrices describing the evolution, we are able to obtain analytical relations between the input and output modes from the Bloch-Messiah decomposition of the Heisenberg-picture propagator.

    %%%%%%%%%%%%%%%
	%Figure 1
	%%%%%%%%%%%%%%%
    %%%%%%%%%%%%%%%
	%Figure 1
    %%%%%%%%%%%%%%%

\begin{figure}[t]
		\includegraphics[width=\linewidth]{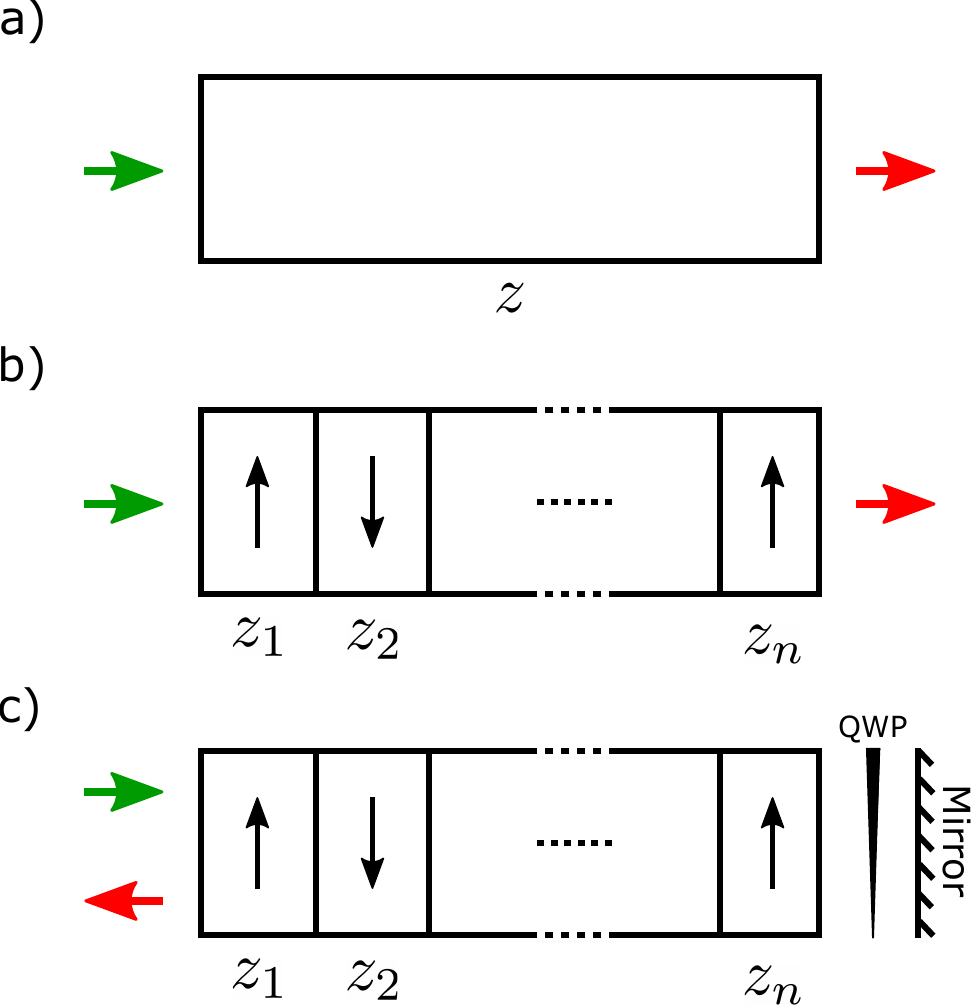}
		\caption{Waveguide poling configurations ($g(z)$)  considered for the Bloch-Messiah decompositions of type-\RomanNumeralCaps{2} SPDC. a) Unpoled single pass structure of length $z$, where the pump pulse enters from the left and the squeezed light exits to the right. b) Poled single pass structure where the nonlinear region is domain engineered(represented by up/down arrows). The domains can be chosen such that $z_{1}=z_{2}=\cdots=z_{n}$ to generate quasi-phase-matching or more arbitrary to generate more exotic phase-matching functions (e.g. Gaussian). c) Poled double pass structure where after the first pass, the beams are reflected back into the sample by reflecting off a mirror. The polarizations of the signal and idler are also swapped by passing through a quarter-wave plate (QWP) twice. Similar to the poled single pass, the values for the $z_{i}$'s can be chosen to generate quasi-phase-matching or more exotic phase-matching functions. In all configurations we assume no reflections at any of the interfaces and no propagation losses.}
		\label{fig:model}
	\end{figure}

    We study three different waveguided squeezers used for squeezed state generation. The first configuration, shown in Fig.~\ref{fig:model}(a), consists of a single squeezer with an unapodized nonlinearity. For the second configuration, shown in Fig.~\ref{fig:model}(b), we consider a single squeezer which has a poled nonlinearity. We consider both quasi-phase-matching (QPM) poling and more arbitrary poling generally used to apodize the phase-matching function~\cite{branczyk2010optimized,dixon2013spectral,tambasco2016domain,dosseva2016shaping}. For the third configuration, shown in Fig.~\ref{fig:model}(c), we consider a double pass structure where after passing through the squeezer once, the beams are passed through a quarter-wave plate (QWP), reflected by a mirror, and then pass through the QWP a second time before going through the squeezer a second time. The double pass through the QWP swaps the polarization of the signal and idler beams and for the second pass, the poling is reversed. This kind of configuration has been used in a number of experiments such as the ones described in Refs.~\onlinecite{zhong2021exp2,lamas2001stimulated,Eisenberg2004doublepass}. Note that we assume that any reflection losses at the interfaces of the squeezers and from the mirror are negligible, as has been the case experimentally\cite{zhong2020exp1, zhong2021exp2}. Non-perfect transmissions between the different elements in these setups could be considered using a more sophisticated formalism such as the one from Ref.~~\onlinecite{Liscindi2012Asymptotic}, however, for the sake of simplicity we consider geometries where reflections are negligible.

    Parametric waveguided squeezers can typically generate squeezed states via two different processes: spontaneous parametric down conversion (SPDC) and four-wave mixing~\cite{quesada2022BPP}. We focus on type-\RomanNumeralCaps{2} SPDC processes where the generated signal and idler photons have degenerate central frequencies but orthogonal polarization. The equations of motion which govern the twin-beam generation are integro-differential equations which, to the best of our knowledge, harbour no analytical solution (except in the limit of a monochromatic pump). We focus on the simplest form of these equations and ignore parasitic phase-matched third-order nonlinear process such as self- and cross-phase modulation. These effects are in general negligible in the limit that we have a large classical pump and a low number of photons in the signal and idler beams~\cite{triginer2020cascaded}. As these effects are ignored, the equations of motion used only focus on $\chi^{2}$ interactions and ignore four-wave mixing processes. Loss is also ignored but could possibly be implemented by building upon the methods developed in this paper. Furthermore, we only focus on the equations of motion governing the mode structure along the direction of propagation and ignore the transverse degrees of freedom. For inline squeezing, there are heralding processes which only focus on the longitudinal modes~\cite{tiedau2019scalability, engelkemeier2021climbing,branczyk2010optimized,blay2017effects,meyer2017limits,thomas2021general}. For quantum computing purposes, one generally wants a squeezer to be as spectrally pure as possible~\cite{vernon2019scalable}. In Ref.~\onlinecite{houde2023sources}, it was found that the double pass structure of Fig.~\ref{fig:model}(c), along with a poling function that gives rise to a Gaussian phase-matching function, a Gaussian pump, and symmetric group-velocity-matching (SGVM) led to optimal spectral purity across a broad range of brightnesses. It remains to be verified if this, or any, configuration gives rise to a perfect inline squeezer.   
    	%%%%%%%%%%%%%%%
	%Figure 2
	%%%%%%%%%%%%%%%
    %%%%%%%%%%%%%%%
	%Figure 2
	%%%%%%%%%%%%%%%
 
    In this paper we focus on non-degenerate waveguided squeezers, however, one could also consider other architectures such as resonators and degenerate squeezers. In Table~\ref{table:degenerate} we summarize known results on the spectral purity and mode-matching of degenerate ring-like and waveguided structures. It has been shown that ring-like structures can be spectrally pure, however, the input and output modes are not matched\cite{quesada2022BPP}, just as ring-like structures that have no nonlinear response. Therefore, ring-like structures are not good candidates for perfect inline squeezers. As for degenerate waveguided squeezers, it has been shown that they are highly spectrally impure when the linear dispersion dominates\cite{Helt2020Degen}. It remains to be seen if such squeezers can be mode-matched, however, simply based on the spectral purity they are not good candidates for perfect inline squeezers. In Table~\ref{table:nondegenerate} we summarize results related to this criteria but for non-degenerate ring-like and waveguided structures. The same results and conclusions hold for non-degenerate ring-like structures. As mentioned previously, non-degenerate waveguided sources can be spectrally pure and, as we show in this paper, can also be mode-matched therefore giving rise to a perfect inline squeezer. 
    Finally, we note that if beamsplitter operations are easy to access between the two beams of the twin-beam (as is the case for frequency degenerate type-\RomanNumeralCaps{2}, for which they correspond to polarizing beamsplitters) then it is easy to show that a perfect inline squeezer can be turned into two single-mode squeezing operations acting in parallel on the two modes~\cite{kalajdzievski2021exact}.

	\begin{table}[!t]
    \begin{tabular}{ |c|c|c| } 
    \hline
     Degenerate Squeezing & Rings & Waveguides\\ \hline \hline
    Spectrally Pure & \ding{51} & \ding{55} \\ \hline
    Input-Output Mode-Matched & \ding{55} & ?  \\ \hline
    \end{tabular}
    \caption{Comparing degenerate squeezing structures. Ring-like structures have been shown to be spectrally pure but not mode-matched \cite{quesada2022BPP} and are therefore not perfect inline squeezers. Degenerate waveguided structures are spectrally impure\cite{Helt2020Degen} and as such are not perfect inline squeezers (to the best of our knowledge no work has been done to study the input/output temporal modes). }
    \label{table:degenerate}
    \end{table}

	\begin{table}[!t]
    \begin{tabular}{ |c|c|c| } 
    \hline
     Twin-Beam Squeezing & Rings & Waveguides\\ \hline \hline
    Spectrally Pure & \ding{51} & \ding{51} \\ \hline
    Input-Output Mode-Matched & \ding{55} & \ding{51}  \\ \hline
    \end{tabular}
    \caption{Comparing twin-beam squeezing structures. Ring-like structures have been shown to be spectrally pure but not mode-matched \cite{quesada2022BPP} and are therefore not perfect inline squeezers. Twin-beam waveguided structures have been shown to be spectrally pure\cite{houde2023sources} and, as we show in this paper, can also be mode-matched. Twin-beam waveguided squeezers can therefore be perfect inline squeezers.}
    \label{table:nondegenerate}
    \end{table}
 
	By first considering the configuration of Fig.~\ref{fig:model}(a) and studying the matrix structure of the discretized equations of motion, we were able to determine an orthogonal transformation which turns the Heisenberg-picture propagator into block-diagonal form. Then, by studying the structure of these blocks, we managed to find an involutory matrix, i.e. a matrix that is its own inverse, which when applied to said blocks gives us a real symmetric matrix. Using the spectral decomposition of real symmetric matrices then allows us to break down the Heisenberg-picture propagator into the form of a Bloch-Messiah decomposition. By directly comparing the input and output modes, we are able to show that they are indeed different. Considering a more specific pump pulse, e.g. Gaussian, we find that the input and output modes are flips of each other in frequency space (relative to their central frequencies), which explains the numerical results shown in Fig.~\ref{fig:modes}(a)(b).

    We then move on to the configuration of Fig.~\ref{fig:model}(b). We first consider poling which gives rise to QPM. In this case, we find that the procedure used for the configuration of Fig.~\ref{fig:model}(a) also gives us the Bloch-Messiah decomposition. As such, the resulting input and output modes differ. We then consider arbitrary poling. Since there are no assumptions made about the poling, we cannot obtain an involutory matrix which makes the matrices symmetric. However, by using the singular-value decomposition of the diagonal blocks, we can still obtain a general Bloch-Messiah decomposition which shows that the input and output modes differ. Unless we specifically tailor the poling to give rise to symmetric diagonal blocks, arbitrary poling will result in differing input and output modes. 
    If the pump pulse is taken to be Gaussian, we can also argue, from the underlying symmetries, that the input and output modes will also have the same flip structure as the non-poled case.   
    
    For the double pass configuration of Fig.~\ref{fig:model}(c), we show that the diagonal blocks are themselves symmetric. We do not need to multiply by an extra involutory matrix. Thus, the input and output modes obtained from the Bloch-Messiah decomposition are the same which explains the numerical results shown in Fig.~\ref{fig:modes}(c)(d). The double pass structure has a desirable property due to the polarization swap: One finds that the input(output) modes of the second pass are the output(input) modes of the first, leading to a configuration which has the same overall input and output modes.

    Finally, we consider all three configurations when we are not in the SGVM regime. In this case, the orthogonal transformation differs as we lose some internal symmetries. We find that all of the configurations have different input and output modes, even the double pass structure. It is thus important to be in the SGVM regime, which consequently is also required to obtain high spectral purity. From our Bloch-Messiah analysis, \emph{we find that the double pass structure, in the SGVM regime, is a perfect inline squeezer}. 
 
	The paper is organized as follows. In Sec.~\ref{sec:Issues} we provide a detailed analysis of why it is important to match input and output modes when squeezing states other than the vacuum. In Sec.~\ref{sec:Bloch-Messiah}, we give a brief description of the Bloch-Messiah decomposition as well as how to identify the input and output Schmidt modes from said decomposition. In Sec.~\ref{sec:Model} we describe the model used for type-\RomanNumeralCaps{2} SPDC as well as the underlying assumptions used in its derivation. In Sec.~\ref{sec:NumericalBM}, we go over numerical results for the Bloch-Messiah decomposition of certain waveguided squeezers (shown in Fig.~\ref{fig:model}) and show the difference (or lack thereof) between the input and output Schmidt modes to motivate the need to understand how these differences arise for different configurations. In Sec.~\ref{sec:AnalyticBM} we develop a method to obtain analytic relations between input and output modes from the Bloch-Messiah decomposition of the discretized equations of motion by using symmetries of the underlying matrix which governs the evolution. We build our solution by first considering the SGVM regime for an unapodized squeezer in Sec.~\ref{sec:nopoling}. We then consider the simplest poling which gives rise to QPM in Sec.~\ref{sec:qpm} and move on to arbitrary poling (i.e. apodized poling to give rise to a Gaussian phase-matching function) in Sec.~\ref{sec:arbpoling}. In Sec.~\ref{sec:doublepass} we consider the double pass structure of Fig.~\ref{fig:model}(c). In Sec.~\ref{app:fidelity}, we numerically consider how having distinct passes with different levels of gain affects the double pass result. In Sec.~\ref{app:nosgvm}, we consider more general conditions relaxing the SGVM constraint. Table \ref{table:assumptions} concisely summarizes which assumptions are made for each section. Finally, in Sec.~\ref{sec:conc} we conclude with our findings.

    \begin{table}[h]
    \begin{tabular}{ |c|c| } 
    \hline

    Section & Assumptions  \\\hline \hline
    \ref{sec:nopoling} & SGVM, real pump, no poling \\ \hline     
    \ref{sec:evenpump} & Same as \ref{sec:nopoling} + frequency symmetric pump  \\ \hline
    \ref{sec:qpm} & SGVM, real pump, QPM poling \\ \hline
    \ref{sec:arbpoling} & SGVM, real pump, arbitrary poling \\ \hline
    \ref{sec:doublepass} & SGVM, real pump, arbitrary poling \\ \hline
    \ref{app:fidelity} & Same as \ref{sec:doublepass}+variable gain(numerical) \\ \hline
    \ref{app:nosgvm} & real frequency symmetric pump, no poling\\
    \hline
    \end{tabular}
    \caption{Assumptions taken for the different sections. SGVM stands for symmetric group velocity matching.}
    \label{table:assumptions}
    \end{table}

	%%%%%%%%%%%%%%%%%%%%%%%%%%%%%%%%%%%%%%%%%%%%%%%%%%%%%%%%%%%%%%%%%%%%%%%%
	%%%%%%%%%%%%%%%%%%%%%%%%%%%%%%%%%%%%%%%%%%%%%%%%%%%%%%%%%%%%%%%%%%%%%%%%
	\section{Issues with inline squeezing}
	\label{sec:Issues}
	%%%%%%%%%%%%%%%%%%%%%%%%%%%%%%%%%%%%%%%%%%%%%%%%%%%%%%%%%%%%%%%%%%%%%%%%
	%%%%%%%%%%%%%%%%%%%%%%%%%%%%%%%%%%%%%%%%%%%%%%%%%%%%%%%%%%%%%%%%%%%%%%%%
    The states generated by two-mode squeezers, such as waveguided squeezers, can be described by the action of an evolution operator associated with a general quadratic Hamiltonian which (up to a global phase) takes the form  
    \begin{align}\label{eq:operator}
        \bm{\mathcal{U}} = \bm{\mathcal{U}}_{s}\bm{\mathcal{U}}_{0},
    \end{align}
    where $\bm{\mathcal{U}}_{0}$ is a unitary operator that satisfies $\bm{\mathcal{U}}_{0}|\text{vac}\rangle~=~|\text{vac}\rangle$ and $\bm{\mathcal{U}}_{s}$ is a unitary operator responsible for the squeezing operation $\bm{\mathcal{U}}_{s}|\text{vac}\rangle=|\text{TMSV}\rangle$ (TMSV meaning two-mode squeezed vacuum)\cite{quesada2022BPP}. In terms of the output Schmidt modes, we can express the squeezing operator as
    \begin{align}
        \bm{\mathcal{U}}_{s} = \text{exp}\left[ \sum_{k} r_{k}A_{k}^\dagger B_{k}^\dagger - \text{H.c.} \right],
    \end{align}
    where the $r_{k}$'s are the squeezing parameters, $A_{k}$($B_{k}$) are the signal(idler) output Schmidt modes, and H.c. is the hermitian conjugate. In terms of the input Schmidt modes, we can express the first unitary as
    \begin{align}
        \bm{\mathcal{U}}_{0} = \text{exp}\left[ -i \sum_{k} \phi_{k,a}\mathcal{A}_{k}^\dagger \mathcal{A}_{k} \right]\text{exp}\left[ -i \sum_{k} \phi_{k,b}\mathcal{B}_{k}^\dagger \mathcal{B}_{k} \right],
    \end{align}
    where the $\phi_{k,a/b}$'s are real values and $\mathcal{A}_{k}$($\mathcal{B}_{k}$) are the signal(idler) input Schmidt modes. In general, focusing on the signal Schmidt modes only, the set of input and output modes are different and can be related to each other by a unitary matrix $U$ as follows
    \begin{align} \label{eqn:unitarytrans}
        A^{\dagger}_{k} &= \sum_{l}U_{k,l}\mathcal{A}^{\dagger}_{l},  \quad 
        \mathcal{A}^{\dagger}_{k} = \sum_{l} U^{*}_{l,k}A^{\dagger}_{l}. 
    \end{align}
    To see why the potential mismatch between input and output can cause issues when doing inline squeezing we consider the following situation: We first use the squeezer to generate a TMSV. We then measure the idler to herald the signal photons in a well defined Schmidt mode, which we assume, for concreteness' sake, to be the first Schmidt mode, such that the output state is a single photon,
    \begin{align}
        |\Psi_{\text{I}}\rangle = A^{\dagger}_{1}|\text{vac}\rangle.
    \end{align}
    We then use this state as a seed for the same squeezer. Note that we could also seed with a more general state described by $f(a^{\dagger}_{S,n}) \ket{\text{vac}}$ for some function of the signal beam operators $a_{S,n}^\dagger$ where $n$ labels other degrees of freedom (i.e. frequency, or momentum). We would now want to (two-mode) squeeze the state $|\Psi_{\text{I}} \rangle$, however, because of the form of the evolution operator in Eq.~\ref{eq:operator}, it is not necessarily as simple as using the same squeezer on the heralded state. Indeed, if one seeds the same squeezer with the heralded state, the final state is
    \begin{align}
        |\Psi_{\text{F}}\rangle&=\bm{\mathcal{U}}|\Psi_{\text{I}}\rangle\nonumber\\
            &=\bm{\mathcal{U}}A^{\dagger}_{1}\bm{\mathcal{U}}^{\dagger}\bm{\mathcal{U}}|\text{vac}\rangle\nonumber\\
            &=\bm{\mathcal{U}}_{s}\left(\bm{\mathcal{U}}_{0}A^{\dagger}_{1}\bm{\mathcal{U}}^{\dagger}_{0}   \right)|\text{vac}\rangle.
    \end{align}
    If the input and output Schmidt modes are equivalent (i.e. $U_{k,l}=\delta_{k,l}$ in Eq.~\ref{eqn:unitarytrans}) then $\bm{\mathcal{U}}_{0}A^{\dagger}_{1}\bm{\mathcal{U}}^{\dagger}_{0}= e^{i \phi_{1,a}} A^{\dagger}_{1}$ and the final state is the desired state (up to a global phase). However, in general we find that 
    \begin{align}
        \bm{\mathcal{U}}_{0}A^{\dagger}_{1}\bm{\mathcal{U}}^{\dagger}_{0} &= \sum_{l,m} U_{1,l}e^{i\phi_{l,a}}U^{*}_{m,l}A^{\dagger}_{m}\nonumber\\
        & = \sum_{m}\left[ \bm{U}e^{i\bm{\Phi}}\bm{U}^{\dagger}  \right]_{1,m}A^{\dagger}_{m},
    \end{align}
    where in the last line we represent the summation over the $l$ index as a matrix multiplication with $e^{i\bm{\Phi}}~\equiv~\text{diag}(e^{i\phi_{1,a}},e^{i\phi_{2,a}},\ldots, e^{i\phi_{n,a}})$. If $e^{i\bm{\Phi}}$ is the identity, which is generally not the case, we again have that $\bm{\mathcal{U}}_{0}A^{\dagger}_{1}\bm{\mathcal{U}}^{\dagger}_{0}= A^{\dagger}_{1}$ and the final state is the desired squeezed state. When $e^{i\bm{\Phi}}$ is anything other than the identity, we see that we have a mixture of the output Schmidt modes and so the final state has many additional unwanted parts. Even if we consider the spectrally pure case, where the squeezing parameters $r_{k}$ and phases $\phi_{k,j}$ are all zero except for $r_{1}$ and $\phi_{1,a}$ which are finite, the final state will have many contributions from the creation operators $A^{\dagger}_{m>1}$, acting on the vacuum as we show below, 
    \begin{align}
        \bm{\mathcal{U}}_{0}A^{\dagger}_{1}\bm{\mathcal{U}}^{\dagger}_{0} &= \sum_{m} \left[ \delta_{1,m}+(e^{i\phi_{1,a}}-1)U_{1,m}U^{*}_{m,1}\right]A^{\dagger}_{m}.
    \end{align}

    We can also see how this effect comes up in the Heisenberg picture by considering the Bloch-Messiah decomposition of the Heisenberg-picture propagator. The Bloch-Messiah decomposition gives us information concerning the input Schmidt modes, responsible for the effects of $\bm{\mathcal{U}}_{0}$, and the output Schmidt modes, responsible for $\bm{\mathcal{U}}_{s}$. Obtaining analytical information concerning the Bloch-Messiah decomposition is thus important in fully characterizing squeezers and determining how to use them optimally to generate squeezed states in a well-defined mode.

    %%%%%%%%%%%%%%%%%%%%%%%%%%%%%%%%%%%%%%%%%%%%%%%%%%%%%%%%%%%%%%%%%%%%%%%%
	%%%%%%%%%%%%%%%%%%%%%%%%%%%%%%%%%%%%%%%%%%%%%%%%%%%%%%%%%%%%%%%%%%%%%%%%
	\section{Bloch-Messiah decomposition}
	\label{sec:Bloch-Messiah}
	%%%%%%%%%%%%%%%%%%%%%%%%%%%%%%%%%%%%%%%%%%%%%%%%%%%%%%%%%%%%%%%%%%%%%%%%
	%%%%%%%%%%%%%%%%%%%%%%%%%%%%%%%%%%%%%%%%%%%%%%%%%%%%%%%%%%%%%%%%%%%%%%%%

    In this section we give a brief overview of the Bloch-Messiah decomposition and how we obtain the Schmidt modes from it. Since we are considering twin-beam squeezing for the rest of the paper, we consider signal($a_{S,n}$) and idler($a_{I,n}$) modes and introduce the hermitian phase quadratures $X_{S/I,n}$ and $P_{S/I,n}$ such that
    \begin{align}
     a_{S/I,n}~=~\tfrac{1}{\sqrt{2\hbar}}\left( X_{S/I,n}+i P_{S/I,n}  \right).
     \end{align} 
 In the definitions in the previous paragraph, $n$ is a discrete label for the frequency (or longitudinal momentum) modes in the waveguides. Note that in general these modes are quasi-continuous, but can be well-approximated by a sufficiently fine lattice of discrete points. 
 We define a row vector for $N$ discrete modes 
 \begin{align}
 \bm{r}^{T}~=~(& X_{S,1},\ldots,X_{S,N}, X_{I,1},\ldots,X_{I,N}\\ & P_{S,1},\ldots,P_{S,N}, P_{I,1},\ldots,P_{I,N}). \nonumber
 \end{align} From this, we can construct the covariance matrix of a zero-mean ($\langle \bm{r} \rangle = 0$) output state generated by a waveguided squeezer as $\bm{V}~=~\langle \{\bm{r},\bm{r}^{T} \}  \rangle/2 $ where $\{A,B  \}~=~AB+BA$ is the anti-commutator. Squeezing is a Gaussian operation and so the output state is fully described by its covariance matrix. Furthermore, for a pure state we can express the covariance matrix as
    \begin{align}
        \bm{V} = \tfrac{\hbar}{2}\bm{S}\bm{S}^{T},
    \end{align}
    where $\bm{S}$ is a symplectic matrix of even dimension~\cite{serafiniBook}. $\bm{S}$ is also the Heisenberg-picture propagator which relates the quadratures at the end of the squeezer to those at the input
    \begin{align}
        \bm{r}^{\text{out}}=\bm{S}\bm{r}^{\text{in}} = \mathcal{U}^\dagger \bm{r}^{\text{in}} \mathcal{U},
    \end{align}
    and so $\bm{S}$ can be obtained by solving the Heisenberg equations of motion relevant to the problem.
    
    Any symplectic matrix can be decomposed as
	\begin{align}\label{eq:BM}
		\bm{S} = \bm{O}\bm{\Lambda} \bm{\tilde{O}}^{T},
	\end{align}
	where $\bm{O}$ and $\bm{\tilde{O}}$ are both orthogonal and symplectic and $\bm{\Lambda}~=~\text{diag}(\lambda_{1},\ldots,\lambda_{2N},\lambda^{-1}_{1},\ldots,\lambda^{-1}_{2N})$. This is known as its Bloch-Messiah decomposition~\cite{serafiniBook}. As we are considering twin-beam squeezing operations, we know that the diagonal matrix takes the form
	\begin{align}
		\bm{\Lambda}= \text{diag}(&e^{r_{1}},e^{r_{1}},\ldots,e^{r_{N}},e^{r_{N}},\nonumber\\
		&e^{-r_{1}},e^{-r_{1}},\ldots,e^{-r_{N}},e^{-r_{N}})
	\end{align}
	which represents a set of pairs of single-mode squeezing operations with squeezing parameters $\{r_{1},\ldots, r_{N}  \}$~\cite{Cariolar2016BM1,Cariolaro2016BM2,mccutcheon2018structure,horoshko2019bloch}. As we are interested in two-mode squeezing, we need to convert the single-mode squeezing operations into two-mode squeezing operations to properly describe the output state.
	
	We can convert two single-mode squeezing operations into a two-mode squeezing operation, by combining them with a $50:50$ symmetric beamsplitter given by
	\begin{align}
		\bm{w} = \tfrac{1}{\sqrt{2}}\begin{pmatrix}
			1 & i \\
			i & 1
		\end{pmatrix}
	\end{align}
	which we can extend to $N$ pairs of single-mode squeezed state as
	\begin{align}
		\bm{B}_{50:50} = \bigoplus_{k=1}^{N}\bm{w}.
	\end{align}
	To transform the $\bm{\Lambda}$ matrix to two-mode squeezing operations, we need to act on it with
	\begin{align}
		\bm{W} = \begin{pmatrix}
			\Re\left(\bm{B}_{50:50} \right) & -\Im\left(\bm{B}_{50:50} \right)\\
			\Im\left(\bm{B}_{50:50} \right) & \Re\left(\bm{B}_{50:50} \right)
		\end{pmatrix}
	\end{align}
    such that
	\begin{align}
		\bm{S} &= \bm{O}\bm{W}\bm{W}^{T}\bm{\Lambda}\bm{W}\bm{W}^{T} \tilde{\bm{O}}^{T}\nonumber\\
        &= \bm{O}\bm{W}\bm{\Lambda}_{\text{TMS}}\left( \tilde{\bm{O}}\bm{W} \right)^{T},
	\end{align}
    where $\bm{\Lambda}_{\text{TMS}}$ represents two-mode squeezing operations.

    The columns of the $\bm{O}\bm{W}$ matrix give information on the output Schmidt modes whereas those of $\tilde{\bm{O}}\bm{W}$ give information on the input Schmidt modes. All symplectic orthogonal matrices can be expressed as
    \begin{align}
        \bm{O} = \begin{pmatrix}
            \Re(\bm{U}) & -\Im(\bm{U})\\
            \Im(\bm{U}) & \Re(\bm{U})\\
        \end{pmatrix},
    \end{align}
    where $\bm{U}$ is a unitary matrix\cite{serafiniBook} whose column vectors represent the Schmidt modes. Therefore, the first half of the columns of $\bm{O}\bm{W}$ and $\tilde{\bm{O}}\bm{W}$ gives the real parts of the Schmidt modes and the second half gives the imaginary parts. 
    In general, these two matrices are distinct and as such the input and output modes are different. This difference is important when one wants to characterize an inline squeezer as highlighted in the previous section.
    
    %%%%%%%%%%%%%%%%%%%%%%%%%%%%%%%%%%%%%%%%%%%%%%%%%%%%%%%%%%%%%%%%%%%%%%%%
	%%%%%%%%%%%%%%%%%%%%%%%%%%%%%%%%%%%%%%%%%%%%%%%%%%%%%%%%%%%%%%%%%%%%%%%%
	\section{Model of twin-beam generation}
	\label{sec:Model}
	%%%%%%%%%%%%%%%%%%%%%%%%%%%%%%%%%%%%%%%%%%%%%%%%%%%%%%%%%%%%%%%%%%%%%%%%
	%%%%%%%%%%%%%%%%%%%%%%%%%%%%%%%%%%%%%%%%%%%%%%%%%%%%%%%%%%%%%%%%%%%%%%%%
	
	We use the twin-beam dynamical equations derived in Ref.~~\onlinecite{quesada2020theory} and used in Ref.~~\onlinecite{houde2023sources}. We briefly describe the underlying assumptions and express the relevant equations of motion.
	
	Without loss of generality, we assume that the beams propagate in the $z$ direction. We assume that each beam, $j=P,S,I$ for pump, signal, and idler respectively, has a linear dispersion around a central frequency $\bar{\omega}_{j} = \omega_{j,\bar{k}_{j}}$, which is associated with a central wavevector $\bar{k}_{j}$, such that 
    \begin{align}\label{Eq:DispersionMT}
		\omega_{j,k} \approx \bar{\omega}_{j} +v_{j}(k-\bar{k}_{j}).
	\end{align}
    The group velocity $v_{j}$ is assumed to be constant over the relevant frequency ranges thus we ignore group velocity dispersion.
    We focus on type-\RomanNumeralCaps{2} SPDC processes and require that
	\begin{align}
		\bar{\omega}_{P}-\bar{\omega}_{S}-\bar{\omega}_{I}&=0,\\
		\bar{k}_{P}-\bar{k}_{S}-\bar{k}_{I}&=0.\label{Eq:MomconMT}
	\end{align}
	For QPM, the right-hand side of Eq.~\ref{Eq:MomconMT} should be changed to $\pm 2\pi/\Lambda_{\text{pol}}$ where $\Lambda_{\text{pol}}$ is the poling period. As in Ref.~~\onlinecite{houde2023sources}, we consider that we are in the undepleted-classical pump regime and assume that self- and cross-phase modulation effects are negligible. These assumptions lead to the following equations for the spatial evolution of the signal and idler operators
	\begin{align}
		\frac{\partial}{\partial z}a_{S}(z,\omega) =& i\Delta k_{S}(\omega)a_{S}(z,\omega)\label{Eq:SigEomZ}\\
		&+i\frac{\gamma g(z)}{\sqrt{2\pi}}\int d\omega' \beta_{P}(z,\omega+\omega')a^{\dagger}_{I}(z,\omega'),\nonumber\\
		\frac{\partial}{\partial z}a^{\dagger}_{I}(z,\omega) =& -i\Delta k_{I}(\omega)a^{\dagger}_{I}(z,\omega)\label{Eq:IdEomZ}\\
		&-i\frac{\gamma^{*} g(z)}{\sqrt{2\pi}}\int d\omega' \beta^{*}_{P}(z,\omega+\omega')a_{S}(z,\omega'),\nonumber
	\end{align}
	where
	\begin{align}
		\Delta k_{S}(\omega)=\left(\frac{1}{v_{S}} - \frac{1}{v_{P}}\right)(\omega_{S}-\bar{\omega}_{I}),\\
        \Delta k_{I}(\omega)=\left(\frac{1}{v_{I}} - \frac{1}{v_{P}}\right)(\omega_{I}-\bar{\omega}_{I}).
	\end{align}
	For the second term on the right-hand side, responsible for twin-beam generation, we have the coupling parameter $\gamma g(z)$ which is related to intrinsic properties of the nonlinearity (see Ref.~~\onlinecite{quesada2020theory,houde2023sources}). The constant $\gamma$ determines the overall strength of the interaction and $g(z)$ is the poling function with $g(z)=0$ where the nonlinearity is absent and either 1 or -1 depending on the orientation of the nonlinear region. Fig.~\ref{fig:model} shows the different orientations we are considering for the poling. The quantity $\beta_{P}(z,\omega)=\beta_{P}(\omega)$, which appears in the integrals, is the spectral content of the pump and under our assumptions is independent of $z$.

	\subsection{Solving the equations of motion}
 
    The equations of motion (Eqs.~\ref{Eq:SigEomZ},~\ref{Eq:IdEomZ}) are complicated integro-differential equations which cannot be solved readily analytically. Numerically, one can solve them by discretizing over a grid of frequencies. In this paper, we are interested in seeing if something can be said analytically based on the matrix structure of the discretized equations of motion. As such, we begin by discretizing the operators $a_{j}(z,\omega)$ in frequency space such that $\omega_{n} = \omega_{0}+n\Delta\omega|^{N}_{1}$ for an $N$-size grid. To simplify notation, we introduce column vectors with entries
	\begin{align}
		\bm{a}_{S,n}(z) = a_{S}(z,\omega_{n}),\\
		\bm{a}^{\dagger}_{I,n}(z) = a^{\dagger}_{I}(z,\omega_{n}),
	\end{align}
	which allows us to re-write the equations of motion (Eqs.~\ref{Eq:SigEomZ}, \ref{Eq:IdEomZ}) in block-matrix form
	\begin{align}
		\frac{\partial}{\partial z}\begin{pmatrix}
			\bm{a}_{S}(z)\\
			\bm{a}_{I}^{\dagger}(z)
		\end{pmatrix} &= i\begin{pmatrix}
			\bm{G}(z) & \bm{F}(z) \\
			-\bm{F}^{\dagger}(z) & -\bm{H}^{\dagger}(z)
		\end{pmatrix}\begin{pmatrix}
			\bm{a}_{S}(z)\\
			\bm{a}_{I}^{\dagger}(z)
		\end{pmatrix},
	\end{align}
	where the matrix elements are given by
	\begin{align}
		\bm{G}_{n,m}(z) &= \Delta k_{S}(\omega_{n})\delta_{n,m}\label{eq:Gmat},\\
		\bm{H}_{n,m}(z) &= \Delta k_{I}(\omega_{n})\delta_{n,m},\\
		\bm{F}_{n,m}(z) &= \frac{\gamma g(z)}{\sqrt{2\pi}}\beta_{P}(\omega_{n}+\omega_{m})\Delta\omega\label{eq:Fmat}.
	\end{align}    
	Note that the matrices $\bm{G}(z)$, $\bm{F}(z)$, and $\bm{H}(z)$ are all symmetric matrices. Furthermore, the only $z$-dependence appears in the $\bm{F}(z)$ matrix and comes from the poling function $g(z)$ which takes the values $\pm 1$ over short intervals of length $\Delta z$, depending on the orientation of the nonlinear region. Within a given interval, the equations of motion are $z$-independent and can be solved via matrix exponentiation. As we are interested in the Bloch-Messiah decomposition, we re-write the equations of motion in the extended quadrature basis where $\bm{a}_{l}(z)~=~ \left(\bm{X}_{l}(z)+i\bm{P}_{l}(z)   \right)/\sqrt{2\hbar}$ for $l=S,I$. We assume $\gamma \beta_{P}(\omega)$ to be real and obtain
    \begin{align}
		\frac{\partial}{\partial z}\begin{pmatrix}
			\bm{X}_{S}(z)\\
			\bm{X}_{I}(z)\\
            \bm{P}_{S}(z)\\
			\bm{P}_{I}(z)
		\end{pmatrix} &= \tilde{\bm{Q}}(z)\begin{pmatrix}
			\bm{X}_{S}(z)\\
			\bm{X}_{I}(z)\\
            \bm{P}_{S}(z)\\
			\bm{P}_{I}(z)
		\end{pmatrix},
	\end{align}
     where
     \begin{align}
        \tilde{\bm{Q}}(z)=\begin{pmatrix}
			0_{N\times N} & 0_{N\times N} & -\bm{G}(z) & \bm{F}(z) \\
			0_{N\times N} & 0_{N\times N} & \bm{F}(z) & -\bm{H}(z) \\
            \bm{G}(z) & \bm{F}(z) & 0_{N\times N} & 0_{N\times N}  \\
            \bm{F}(z) & \bm{H}(z) & 0_{N\times N} & 0_{N\times N}  \\
		\end{pmatrix}.
     \end{align}
    For a given interval with known value of $g(z)$, the solution is given by
	\begin{align}
		\begin{pmatrix}
			\bm{X}_{S}(z_{0}+\Delta z)\\
			\bm{X}_{I}(z_{0}+\Delta z)\\
   		\bm{P}_{S}(z_{0}+\Delta z)\\
			\bm{P}_{I}(z_{0}+\Delta z)
		\end{pmatrix}= \bm{S}(z_{0}+\Delta z,z_{0})\begin{pmatrix}
			\bm{X}_{S}(z_{0})\\
			\bm{X}_{I}(z_{0})\\
   		\bm{P}_{S}(z_{0})\\
			\bm{P}_{I}(z_{0})\end{pmatrix}.\label{Eq:Soln}
	\end{align}
	where $\bm{S}(z_{0}+\Delta z,z_{0})=\text{exp}\left(\Delta z \tilde{\bm{Q}}(z_{0})\right)$ is the Heisenberg-picture propagator and it is understood that $g(z_{0})$ in $\tilde{\bm{Q}}(z_{0})$ takes the given value over the short interval spanning $\left( z_{0},z_{0} +\Delta z \right)$.

    %%%%%%%%%%%%%%%
	%Figure 3
	%%%%%%%%%%%%%%%
	\begin{figure*}[t]
		\includegraphics[width=0.75\linewidth]{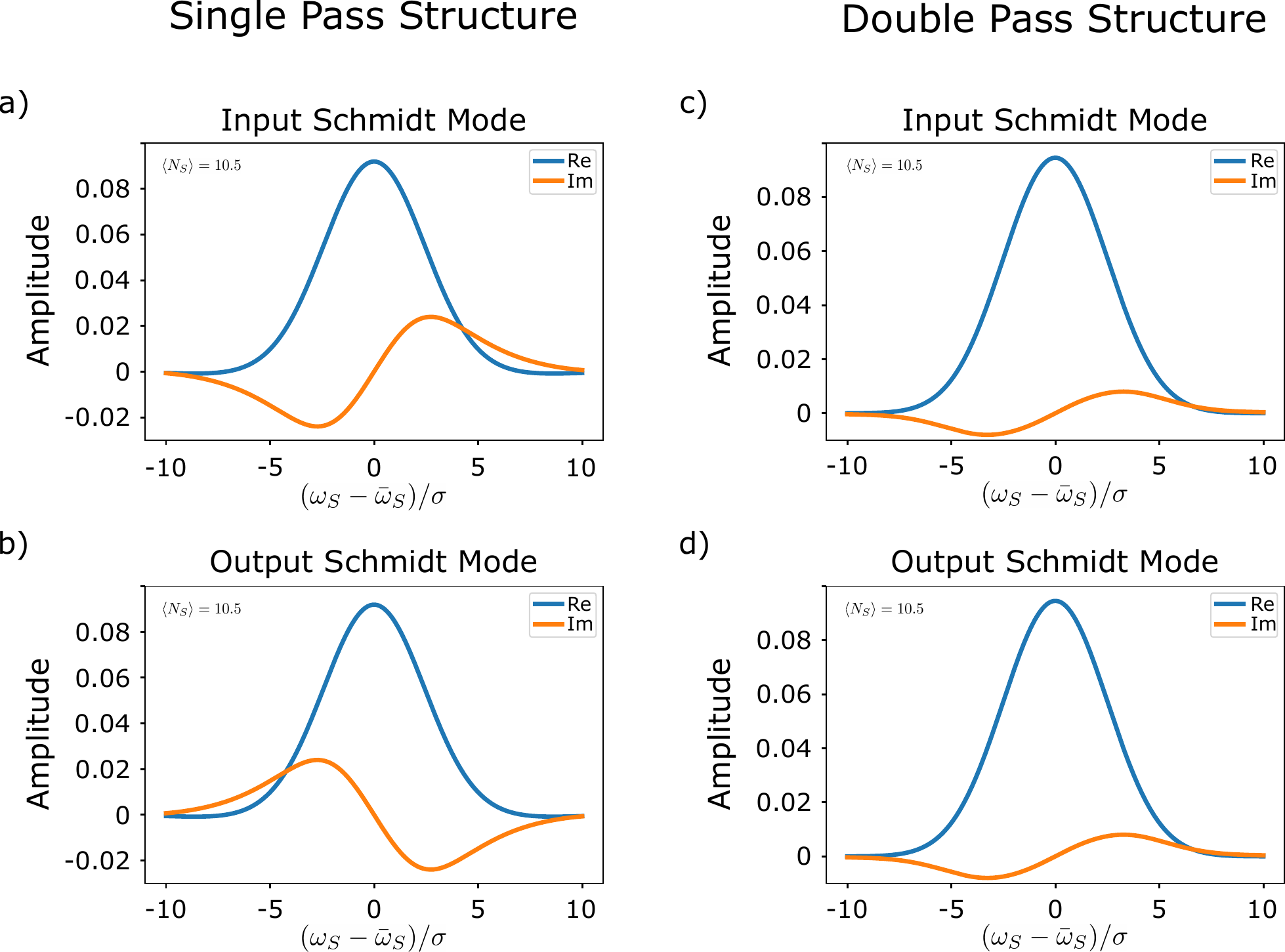}
		\caption{Temporal mode structure obtained from numerical Bloch-Messiah decomposition. Parameters for both structures were chosen to give the same average number of signal photons $\langle N_{S}\rangle = 10.5 $, giving rise to different Schmidt modes. Plots are shown for poling function which gives rise to a Gaussian phase-matching function, however, the same phenomenological results are obtained for unpoled and periodic poling as well as for arbitrary levels of gain. a) First input Schmidt mode for single pass structures. b) First output Schmidt mode for single pass structures. We see that the input and output modes for the single pass structures have opposite phases and are therefore different. c) First input Schmidt mode for the double pass structure. d) First output Schmidt mode for double pass structure. We see that for the double pass structure the phases remain the same and so the input and output Schmidt modes are the same.}
		\label{fig:modes}
	\end{figure*}
    %%%%%%%%%%%%%%%
	%Figure 3
    %%%%%%%%%%%%%%%
 
	For a known poling function, $g(z)$, we can form the complete propagator by chronologically stitching together the solutions over every interval $\Delta z$. The formal solution is given by
	\begin{align}
		\bm{S}(z,z_{0})=\prod_{p}\text{exp}\left(i\Delta z_{p}\tilde{\bm{Q}}(z_{p})    \right),
	\end{align}
	where different segments can possibly be of different lengths.  We now wish to use this matrix structure to obtain analytical relations between the input and output modes of the Bloch-Messiah decomposition of the Heisenberg-picture propagator but before doing so, we briefly discuss numerical results.

    %%%%%%%%%%%%%%%%%%%%%%%%%%%%%%%%%%%%%%%%%%%%%%%%%%%%%%%%%%%%%%%%%%%%%%%%%%%%%%%%%%%%%%%
	\subsection{Numerical Bloch-Messiah results}
	\label{sec:NumericalBM}
	%%%%%%%%%%%%%%%%%%%%%%%%%%%%%%%%%%%%%%%%%%%%%%%%%%%%%%%%%%%%%%%%%%%%%%%%

    Numerically, one can implement the Bloch-Messiah decomposition and study the input and output Schmidt modes for a given set of parameters. Reference~\cite{houde2024decomp} goes through the numerical implementation in details which is included in the Python library \texttt{TheWalrus}~\cite{thewalrus,Gupt2019}. We first begin by obtaining the solutions for the propagator using the Python library \texttt{NeedALight}\cite{needalight,houde2023sources} and study its Bloch-Messiah decomposition. 
    
    We consider both single and double pass structures where the nonlinear regions allow for type-\RomanNumeralCaps{2} SPDC processes where the generated signal and idler photons have degenerate frequencies but orthogonal polarization. The nonlinear regions are poled to give rise to a Gaussian phase-matching function in the low gain regime. This is done following the methods of ~\cite{tambasco2016domain,Graffitti_2017,Aggie2021domain}. Using the Python library \texttt{Custom-Poling}~\cite{custompoling} we break the nonlinear regions into $N_{z} = 1000$ domains to obtain the $g(z)$ which gives an approximate Gaussian phase-matching function. We assume a Gaussian pump profile with spectral content 
	\begin{align}\label{Eq:pumpform}
		\beta_{P}(\omega) = \frac{\sqrt{\hbar \bar{\omega}_{P} N_{P}}}{\sqrt[4]{\pi\sigma^{2}}}\exp\left[-\frac{(\omega-\bar{\omega}_{P})^{2}}{2\sigma^{2}}\right],
	\end{align}    
    where $\sigma$ is the mean bandwidth and $N_{P}$ is the mean number of pump photons. We also work in the SGVM regime where 
	\begin{align}\label{Eq:symgrpvel}
		\kappa\equiv\left( \frac{1}{v_{S}}-\frac{1}{v_{P}}  \right) = -\left( \frac{1}{v_{I}}-\frac{1}{v_{P}}  \right).
	\end{align}
    We then break the frequencies into a grid of $N=501$. For the double pass structure, the nonlinear regions are of the same length and the polarization swap is implemented by swapping the signal and idler velocities. For experimental purposes, the numerics can correspond to a pump with a central wavelength of 776 nm, and degenerate central frequencies of the signal and idler beams of 1552 nm, and a duration of ~200 fs. We also tune the interaction strengths to generate $\langle N_{S}\rangle=\langle N_{I}\rangle=10.5$ photons in both configurations for signal and idler beams.

    Figure~\ref{fig:modes} shows the different input and output modes for both single and double pass structures. Note that as outlined in\cite{quesada2020theory}, we gauge away the free propagating phases and further choose a gauge where the imaginary part of the Schmidt modes is zero at the origin to produce these figures. In this gauge, we see that the input and output modes for the single pass configuration are not the same. Indeed, their phase structures are opposites. This difference is also seen for a single pass structure with no poling or QPM poling but has been left out for brevity. From this, we gather that a single pass structure does not give a perfect inline squeezer. The double pass structure, on the other hand, gives rise to input and output modes that are identical and is thus a perfect inline squeezer. Again these results hold for a double pass structure with no poling or QPM. 

    We now wish to understand why this is the case and how this arises from the underlying matrix structure.

    %%%%%%%%%%%%%%%%%%%%%%%%%%%%%%%%%%%%%%%%%%%%%%%%%%%%%%%%%%%%%%%%%%%%%%%%
	%%%%%%%%%%%%%%%%%%%%%%%%%%%%%%%%%%%%%%%%%%%%%%%%%%%%%%%%%%%%%%%%%%%%%%%%
	\section{Analytic Properties of the Bloch-Messiah Decomposition}
	\label{sec:AnalyticBM}
	%%%%%%%%%%%%%%%%%%%%%%%%%%%%%%%%%%%%%%%%%%%%%%%%%%%%%%%%%%%%%%%%%%%%%%%%
	%%%%%%%%%%%%%%%%%%%%%%%%%%%%%%%%%%%%%%%%%%%%%%%%%%%%%%%%%%%%%%%%%%%%%%%%	
	
    As in the numerical case, we begin by focusing on the SGVM regime. The initial first step in the derivation for the arbitrary velocity case is slightly more convoluted and some of the results that follow apply by similar arguments presented in the matched regime. As such, we come back to the arbitrary velocity case at the end. Again, we also assume type-\RomanNumeralCaps{2} SPDC where signal and idler photons have degenerate frequencies but opposite polarization. In this case, we have that $\bm{G}(z)~=~-\bm{H}(z)$ and the equations of motion simplify.

    %%%%%%%%%%%%%%%%%%%%%%%%%%%%%%%%%%%%%%%%%%%%%%%%%%%%%%%%%%%%%%%%%%%%%%%%
	%%%%%%%%%%%%%%%%%%%%%%%%%%%%%%%%%%%%%%%%%%%%%%%%%%%%%%%%%%%%%%%%%%%%%%%%
	\subsection{Single pass without poling}
	\label{sec:nopoling}
	%%%%%%%%%%%%%%%%%%%%%%%%%%%%%%%%%%%%%%%%%%%%%%%%%%%%%%%%%%%%%%%%%%%%%%%%
	%%%%%%%%%%%%%%%%%%%%%%%%%%%%%%%%%%%%%%%%%%%%%%%%%%%%%%%%%%%%%%%%%%%%%%%%	
    To help build up the solutions, we begin by considering one interval of length $\Delta z$ along which the poling function $g(z)=1$ which is visually represented in Fig.~\ref{fig:model}(a). Under these conditions, the equations of motions are governed by the $z$-independent matrix
    \begin{align}
        \tilde{\bm{Q}}=\begin{pmatrix}
			0_{N\times N} & 0_{N\times N} & -\bm{G} & \bm{F} \\
			0_{N\times N} & 0_{N\times N} & \bm{F} & \bm{G} \\
            \bm{G} & \bm{F} & 0_{N\times N} & 0_{N\times N}  \\
            \bm{F} & -\bm{G} & 0_{N\times N} & 0_{N\times N}  \\
		\end{pmatrix},
    \end{align}        
    where now $\bm{G}_{n,m}~=~\kappa\left(\omega_{n}-\bar{\omega}\right)\delta_{n,m}$ with $\bar{\omega}\equiv\bar{\omega}_{S}=\bar{\omega}_{I}$ and  $\bm{F}_{n,m}~=~\frac{\gamma}{\sqrt{2\pi}}\beta_{P}(\omega_{n}+\omega_{m})\Delta\omega $. The solution to the propagator is $\bm{S}~=~\text{exp}\left( \Delta z \tilde{\bm{Q}} \right)$. In general, this matrix exponential cannot be evaluated analytically and as such we cannot obtain an analytical relation between the input and output modes from the Bloch-Messiah decomposition. However, with certain manipulations and using matrix properties, it is possible to obtain a form which readily factors into a Bloch-Messiah decomposition. 

    We begin by considering the orthogonal matrix
    \begin{align}\label{eq:Bmat}
        \bm{B} = \frac{1}{\sqrt{2}}\begin{pmatrix}
            \mathbb{1}_{N\times N} & 0_{N\times N} & 0_{N\times N} & \mathbb{1}_{N\times N} \\
            0_{N\times N} & \mathbb{1}_{N\times N} & \mathbb{1}_{N\times N} & 0_{N\times N} \\
            0_{N\times N} & -\mathbb{1}_{N\times N} & \mathbb{1}_{N\times N} & 0_{N\times N} \\
            -\mathbb{1}_{N\times N} & 0_{N\times N} & 0_{N\times N} &  \mathbb{1}_{N\times N} \\
        \end{pmatrix},        
    \end{align}
    where $\mathbb{1}_{N\times N}$ is the identity matrix. Since this matrix is orthogonal we can re-write the propagator as
    \begin{align}
        \bm{S} = e^{\Delta z \tilde{\bm{Q}}}=\bm{B}\bm{B}^{T} e^{\Delta z \tilde{\bm{Q}}}\bm{B}\bm{B}^{T}=\bm{B} e^{\Delta z \bm{B}^{T}\tilde{\bm{Q}}\bm{B}}\bm{B}^{T},
    \end{align}
    where we now need to exponentiate the matrix
    \begin{align}
        \bm{\mathcal{Q}}\equiv \bm{B}^{T}\tilde{\bm{Q}}\bm{B}=\begin{pmatrix}
			-\bm{F} & \bm{G} & 0_{N\times N} & 0_{N\times N} \\
			-\bm{G} & -\bm{F} & 0_{N\times N} & 0_{N\times N} \\
            0_{N\times N} & 0_{N\times N} & \bm{F} & \bm{G}  \\
            0_{N\times N} & 0_{N\times N} & -\bm{G} & \bm{F}  \\
		\end{pmatrix}.
    \end{align}
    Being a block-diagonal matrix simplifies the matrix exponentiation as we now need to exponentiate two smaller matrices. We can further simplify this expression by using the fact that both $\bm{F}$ and $\bm{G}$ are symmetric and define a new matrix 
    \begin{align}\label{eq:Amat}
        \bm{A} = \begin{pmatrix}
        -\bm{F} & \bm{G} \\
        -\bm{G}  & -\bm{F}\\
    \end{pmatrix},
    \end{align}
    which gives us
    \begin{align}
        \bm{\mathcal{Q}} = \begin{pmatrix}
        \bm{A} & 0_{2N\times 2N} \\
        0_{2N\times 2N}  & -\bm{A}^{T}\\
    \end{pmatrix}.
    \end{align}
    We have now reduced the problem to finding $\exp \left( \Delta z \bm{A} \right)$. The matrix $\bm{A}$ is itself not symmetric and so there is no nice way to decompose its exponential. We can however consider an extended Pauli X-matrix
    \begin{align}
        \bm{X} &=\begin{pmatrix}
            0_{N\times N} & \mathbb{1}_{N\times N}\\
            \mathbb{1}_{N\times N} & 0_{N\times N}\\ 
        \end{pmatrix}
    \end{align}
    and from the structure of $\bm{A}$ we find that 
    \begin{align}
        \bm{X}\bm{A}\bm{X} &= \bm{A}^{T}
    \end{align}
    and similarly
    \begin{align}\label{eq:Xtransform}
        \bm{X}e^{\Delta z\bm{A}}\bm{X} &= e^{\Delta z\bm{A}^{T}} =\left( e^{\Delta z\bm{A}} \right)^{T}.
    \end{align}
    Multiplying Eq.~\ref{eq:Xtransform} on the right by $\bm{X}$ allows us to show that 
    \begin{align}
        \bm{X}e^{\Delta z\bm{A}} = \left( \bm{X}e^{\Delta z\bm{A}}\right)^{T}.
    \end{align}
    The product $\bm{X}e^{\Delta z\bm{A}}$ is thus a real-symmetric matrix. Following the same arguments, we also find that $\bm{X}e^{-\Delta z\bm{A}^{T}}$ is also a real-symmetric matrix. Using the fact that $\bm{X}^{2}=\mathbb{1}_{2N\times 2N}$, we can express the propagator as
    \begin{align}
    \bm{S} &= \bm{B}    \begin{pmatrix}
        e^{\Delta z\bm{A}} & 0_{2N\times 2N}  \\
         0_{2N\times 2N} & e^{-\Delta z\bm{A}^{T}}  \\
    \end{pmatrix} \bm{B}^{T} \nonumber\\
    &=\bm{B} \begin{pmatrix}
        \bm{X} & 0_{2N\times 2N}  \\
         0_{2N\times 2N} & \bm{X}  \\
    \end{pmatrix}\begin{pmatrix}
        \bm{X}e^{\Delta z\bm{A}} & 0_{2N\times 2N}  \\
         0_{2N\times 2N} & \bm{X}e^{-\Delta z\bm{A}^{T}}  \\
    \end{pmatrix} \bm{B}^{T}. 
    \end{align}
    Now, it is well-known that real-symmetric matrices have eigen-decompositions (as they are normal) and thus we can express
    \begin{align}\label{eq:EigenDecomp}
        \bm{X}e^{\Delta z\bm{A}} = \bm{\Gamma} \bm{\Lambda} \bm{\Gamma}^{T},
    \end{align}
    where $\bm{\Gamma}$ is real and orthogonal and $\bm{\Lambda}$ is real and diagonal. Similarly we have that $\bm{X}e^{-\bm{A}^{T}} = \bm{\tilde{\Gamma}} \bm{\tilde{\Lambda}} \bm{\tilde{\Gamma}}^{T}$ for some $\bm{\tilde{\Gamma}}$ real and orthogonal and $\bm{\tilde{\Lambda}}$ real and diagonal. To end up with a Bloch-Messiah structure, we need to relate the eigenvectors and eigenvalues of both decompositions. This can be done by simply taking the inverse-transpose of Eq.~\ref{eq:EigenDecomp}. Doing so we find that we can decompose
    \begin{align}
        \bm{X}e^{-\Delta z\bm{A}^{T}} = \bm{\Gamma} \bm{\Lambda}^{-1} \bm{\Gamma}^{T}
    \end{align}    
    which tells us that the eigenvectors of $\bm{X}e^{-\Delta z\bm{A}^{T}}$ are the same as those of $\bm{X}e^{\Delta z\bm{A}}$ but with inverse eigenvalues. The propagator can now be expressed as
    \begin{align}
    \bm{S}&=\bm{B} \begin{pmatrix}
        \bm{X}\bm{\Gamma} & \bm{0}  \\
         \bm{0} & \bm{X}\bm{\Gamma}  \\
    \end{pmatrix}\begin{pmatrix}
        \bm{\Lambda} & \bm{0}  \\
         \bm{0} & \bm{\Lambda}^{-1}  \\
    \end{pmatrix} \begin{pmatrix}
        \bm{\Gamma}^{T} & \bm{0}  \\
         \bm{0} & \bm{\Gamma}^{T}  \\
    \end{pmatrix}\bm{B}^{T}, 
    \end{align}
    where $\bm{0}=0_{2N\times 2N}$, which is highly reminiscent of a Bloch-Messiah decomposition. There are still a couple issues that need to be addressed before we can say that this is indeed a proper Bloch-Messiah decomposition as given by Eq.~\ref{eq:BM}. The diagonal matrix has the proper form where each eigenvalue is accompanied by its inverse, however, we need to show that the values are degenerate (as they should for a twin-beam squeezer). To see this, consider the $2N\times2N$ symplectic form
    \begin{align}
        \bm{\Omega}=\begin{pmatrix}
            0_{N\times N} & \mathbb{1}_{N\times N}\\
            -\mathbb{1}_{N\times N} & 0_{N\times N}\\ 
        \end{pmatrix}.
    \end{align}
    Again from the structure of $\bm{A}$, we can show that $\bm{\Omega}\bm{A}\bm{\Omega}^{T}=\bm{A}$ and similarly $\bm{\Omega}e^{\Delta z\bm{A}}\bm{\Omega}^{T}=e^{\Delta z\bm{A}}$. We also have that $\bm{\Omega}\bm{X}\bm{\Omega}^{T}=-\bm{X}$. Now, suppose we have a column eigenvector, $\bm{v}$ of $\bm{X}e^{\Delta z\bm{A}}$ with eigenvalue $\lambda$, then
    \begin{align}
        \bm{X}e^{\Delta z\bm{A}}\bm{v} &= \lambda \bm{v}\nonumber\\
        \Rightarrow \bm{\Omega}\bm{X}\bm{\Omega}^{T}\bm{\Omega}e^{\Delta z\bm{A}}\bm{\Omega}^{T}\bm{\Omega} \bm{v} &= \lambda \bm{\Omega} \bm{v}\nonumber\\
        \Rightarrow -\bm{X}e^{\Delta z\bm{A}}\bm{\Omega} \bm{v} &= \lambda \bm{\Omega} \bm{v}\nonumber\\
        \Rightarrow \bm{X}e^{\Delta z\bm{A}}\bm{\Omega} \bm{v} &= -\lambda \bm{\Omega} \bm{v}.
    \end{align}
    In the second line we have multiplied on the left by $\bm{\Omega}$ and have inserted a few identities. Therefore, for every column eigenvector $\bm{v}$ we also have another column eigenvector $\bm{\Omega}\bm{v}$ with eigenvalue $-\lambda$. From this, we can express the diagonal matrix as $\bm{\Lambda}~=~\text{diag}(\lambda_{1},-\lambda_{1},\lambda_{2},-\lambda_{2},\ldots,\lambda_{N},-\lambda_{N})$ and the orthogonal matrix in column form $\bm{\Gamma}~=~[\bm{v}_{1}|\bm{\Omega}\bm{v}|\bm{v}_{2}| \bm{\Omega}\bm{v}_{2}|\ldots| \bm{v}_{N}|\bm{\Omega}\bm{v}_{N}  ]$. Since we expect the diagonal matrix to be positive, we can bring the negative sign to either side of the diagonal by considering the matrix
    \begin{align}
        \bm{\Sigma}_{z} = \bigoplus_{n=1}^{N}\begin{pmatrix}
        1 & 0  \\
         0 & -1  \\
    \end{pmatrix}
    \end{align}
    which is a block-diagonal matrix composed of $N$ Pauli-Z matrices. Combining everything together, we find that the final form for the Heisenberg-picture propagator is
    \begin{align}
        \bm{S}&=\bm{B} \begin{pmatrix}
            \bm{X}\bm{\Gamma} & \bm{0}  \\
             \bm{0} & \bm{X}\bm{\Gamma}  \\
        \end{pmatrix}\bm{\Sigma}_{z}\begin{pmatrix}
            \left|\bm{\Lambda}\right| & \bm{0}  \\
            \bm{0} & \left|\bm{\Lambda}^{-1}\right|  \\
        \end{pmatrix} \begin{pmatrix}
            \bm{\Gamma}^{T} & \bm{0}  \\
            \bm{0} & \bm{\Gamma}^{T}  \\
        \end{pmatrix}\bm{B}^{T}, 
    \end{align}
    where again $\bm{0}=0_{2N \times 2N}$ and we can now identify
    \begin{align}
        \bm{O} &=\bm{B}\begin{pmatrix}
            \bm{X}\bm{\Gamma} & 0_{2N\times 2N}  \\
            0_{2N\times 2N} & \bm{X}\bm{\Gamma}  \\
        \end{pmatrix}\bm{\Sigma}_{z}\bm{W},\\
        \tilde{\bm{O}} &=\bm{B}\begin{pmatrix}
            \bm{\Gamma} & 0_{2N\times 2N}  \\
            0_{2N\times 2N} & \bm{\Gamma}  \\
        \end{pmatrix}\bm{W},
    \end{align}
    where we have reintroduced $\bm{W}$ to obtain two-mode squeezing as discussed in Sec.~\ref{sec:Bloch-Messiah}. As a final step, we need to verify that $\bm{O}$ and $\tilde{\bm{O}}$ are symplectic and orthogonal. Since all the individual matrices are orthogonal, it is trivial to show that this condition is met. For a general orthogonal matrix, $\bm{M}$, we can re-write the symplectic condition, $\bm{M}^{T}\bm{\Omega}\bm{M}=\bm{\Omega}$, as $\bm{\Omega}\bm{M}\bm{\Omega}^{T}=\bm{M}$. In this form, it is trivial to show that both $\bm{B}$, $\bm{\Sigma}_{z}$, and $\bm{W}$ are symplectic as well as the last remaining contributions in both $\bm{O}$ and $\tilde{\bm{O}}$. 
    
    We have therefore found the Bloch-Messiah decomposition for the Heisenberg-picture propagator when we take $g(z)=1$ and we can clearly see that the input and output modes are different. This therefore tells us that we expect a single pass case with no poling to have separate input and output modes. In fact, because of the matrices involved, we can introduce column eigenvectors of the form $\bm{v}_{i} = (\bm{a}_{i},\bm{b}_{i})^{T}$, where $\bm{a}_{i}$ and $\bm{b}_{i}$ are sub-vectors of length $N$, to further see how different the input and output modes are. Expressing the $\bm{\Gamma}$ matrix in terms of these eigenvectors we have that
    \begin{align}
        \bm{\Gamma} = \left[\begin{matrix}
			 \bm{a}_{1}\\
			 \bm{b}_{1} 
		  \end{matrix} \middle|\begin{matrix}
			 \bm{b}_{1}\\
			 -\bm{a}_{1} 
		  \end{matrix}\middle|\cdots \middle| \begin{matrix}
			 \bm{a}_{n}\\
			 \bm{b}_{n} 
		  \end{matrix} \middle|\begin{matrix}
			 \bm{b}_{n}\\
			 -\bm{a}_{n} 
		  \end{matrix}       \right].
    \end{align}
    Carrying out the matrix multiplication we then find that the symplectic orthogonal matrices are
    \begin{align}
        \bm{O} = \begin{pmatrix}
            0_{N\times 1} & \bm{a}_{1} &  & 0_{N\times 1} & -\bm{b}_{1} &  \\
            \bm{a}_{1} & 0_{N\times 1} &\ldots & \bm{b}_{1} & 0_{N\times 1} &\ldots \\
            0_{N\times 1} & \bm{b}_{1} &  & 0_{N\times 1} & \bm{a}_{1} &  \\
            -\bm{b}_{1} & 0_{N\times 1} &  & \bm{a}_{1} & 0_{N\times 1} &  \\
        \end{pmatrix},
    \end{align}
    \begin{align}
        \tilde{\bm{O}} = \begin{pmatrix}
            0_{N\times 1} & \bm{b}_{1} &  & 0_{N\times 1} & -\bm{a}_{1} &  \\
            \bm{b}_{1} & 0_{N\times 1} &\ldots & \bm{a}_{1} & 0_{N\times 1} &\ldots \\
            0_{N\times 1} & \bm{a}_{1} &  & 0_{N\times 1} & \bm{b}_{1} &  \\
            -\bm{a}_{1} & 0_{N\times 1} &  & \bm{b}_{1} & 0_{N\times 1} &  \\
        \end{pmatrix}.
    \end{align}
    Note that we have only focused on the first pair of eigenvectors. In this form we can readily identify the real and imaginary parts of both the input and output Schmidt modes. The first input Schmidt mode is given by $\bm{b}_{1}-i\cdot\bm{a}_{1}$ whereas the first output Schmidt mode is given by $\bm{a}_{1}-i\cdot\bm{b}_{1}$. 
    
    We see that the input and output modes swap their real and imaginary parts. For the input and output Schmidt modes to be the same, we require $\bm{a}_{i}=\bm{b}_{i}$ which is equivalent to requiring $\bm{X}\bm{\Gamma}=\bm{\Gamma}$. The latter only holds if $ \bm{X}e^{\Delta z\bm{A}}$ is invariant under a similarity transformation by $\bm{X}$ (i.e. $\bm{X}e^{\Delta z\bm{A}}= \bm{X}\bm{X}e^{\Delta z\bm{A}}\bm{X}$). This only occurs when $\bm{G}=0$ which is a very specific case where all three modes propagate at the same velocity ($v_{S}=v_{I}=v_{P}$).
    
    Before we move on to different poling configurations, we consider a more specific form of $\bm{F}$ via the spectral content of the pump $\beta_{P}(\omega)$ to obtain more constrained eigenvectors which show the behaviours observed numerically in Fig.~\ref{fig:modes}.

    %%%%%%%%%%%%%%%%%%%%%%%%%%%%%%%%%%%%%%%%%%%%%%%%%%%%%%%%%%%%%%%%%%%%%%%%
	%%%%%%%%%%%%%%%%%%%%%%%%%%%%%%%%%%%%%%%%%%%%%%%%%%%%%%%%%%%%%%%%%%%%%%%%
	\subsubsection{Frequency symmetric pump}\label{sec:evenpump}
	%%%%%%%%%%%%%%%%%%%%%%%%%%%%%%%%%%%%%%%%%%%%%%%%%%%%%%%%%%%%%%%%%%%%%%%%
	%%%%%%%%%%%%%%%%%%%%%%%%%%%%%%%%%%%%%%%%%%%%%%%%%%%%%%%%%%%%%%%%%%%%%%%%

    To generate the modes presented in Fig.~\ref{fig:modes}, a Gaussian pump pulse was used. In this section, we consider a more general pump pulse which is frequency symmetric around the pump central frequency, $\beta_{P}(\omega)=\beta_{P}(|\omega-\bar{\omega}_{P}|)$, which includes Gaussian pulses. With this extra constraint the $\bm{F}$ matrix (Eq.~\ref{eq:Fmat}) has an additional symmetry which also transfers to the $\bm{A}$ matrix (Eq.~\ref{eq:Amat}).

    To see this new symmetry, recalling that we assume type-\RomanNumeralCaps{2} SPDC where signal and idler photons have degenerate frequencies but opposite polarization and that the poling function $g(z)=1$, we note that
    \begin{align}
        \bm{F}_{n,m} = \frac{\gamma}{\sqrt{2\pi}}\beta_{P}(|(\omega_{n}-\bar{\omega}) +(\omega_{m}-\bar{\omega})  |),
    \end{align}
    where we take $\omega_{n/m}\in (\bar{\omega}-\delta,\bar{\omega}+\delta)$ to span the same region. As mentioned before $\bm{F}$ is symmetric. Now, because we are also frequency symmetric, terms where $(\omega_{n}-\bar{\omega}) +(\omega_{m}-\bar{\omega}) =\pm \alpha$, for some value $\alpha$, will have the same matrix element. Since we are breaking the frequencies into the same $N$-size grids the pairs of indices $(n,m)$ and $(N-m+1,N-n+1)$ lead to the same absolute value and therefore matrix element. The condition that $\bm{F}_{n,m}=\bm{F}_{N-m+1,N-n+1}$ tells us that $\bm{F}$ is persymmetric. We can also define the persymmetric condition by how the matrix is acted on by the exchange matrix $\bm{J}$. The exchange matrix has the elements on the antidiagonal equal to 1 and all others equal to 0. For the $3\times3$ case, $\bm{J}$ takes the form
    \begin{align}
        \bm{J} = \begin{pmatrix}
            0 & 0 & 1 \\
            0 & 1 & 0  \\
            1 & 0 & 0  \\
        \end{pmatrix}.
    \end{align}
    Using the exchange matrix, persymmetric matrices, $\bm{P}$, are defined such that $\bm{P}\bm{J}=\bm{J}\bm{P}^{T}$. In the case where the matrix is also symmetric, as is the case for $\bm{F}$, the matrix is called centrosymmetric. Thus, we have that $\bm{J}_{N\times N}\bm{F} \bm{J}_{N\times N}=\bm{F}$. Note that a matrix itself need not be symmetric to be centrosymmetric but namely have the latter property. One can readily show that $\bm{J}_{N\times N}\bm{G}\bm{J}_{N\times N}=-\bm{G}$ and so the matrix $\bm{G}$ is anti-centrosymmetric. To see how the matrix $\bm{A}$ transforms under the exchange matrix, we express the $2N\times2N$ exchange matrix as being built of two antidiagonal blocks of the $N\times N$ one
    \begin{align}
        \bm{J}_{2N\times 2N} = \begin{pmatrix}
            0_{N\times N} & \bm{J}_{N\times N}  \\
            \bm{J}_{N\times N} & 0_{N\times N}   \\
        \end{pmatrix}.
    \end{align}    
    Applying the transformation, we have that
    \begin{align}
        \bm{J}_{2N\times 2N}\bm{A}\bm{J}_{2N\times 2N}&=\begin{pmatrix}
            -\bm{J}_{N\times N}\bm{F}\bm{J}_{N\times N} & -\bm{J}_{N\times N}\bm{G}\bm{J}_{N\times N}  \\
            \bm{J}_{N\times N}\bm{G}\bm{J}_{N\times N} & -\bm{J}_{N\times N}\bm{F}\bm{J}_{N\times N}   \\
        \end{pmatrix}\nonumber\\ &=\begin{pmatrix}
            -\bm{F} & \bm{G}  \\
            -\bm{G} & -\bm{F}   \\
        \end{pmatrix}=\bm{A}.
    \end{align}    
    The matrix $\bm{A}$ is also centrosymmetric. Since $\bm{X}$ is also centrosymmetric, we also find that $\bm{X}e^{\Delta z \bm{A}}$ is centrosymmetric. Centrosymmetric matrices have been greatly studied and it is known that the eigenvectors come in symmetric and antisymmetric pairs\cite{CANTONI1976centro}. That is to say, if $\bm{v}$ is an eigenvector then it must satisfy $\bm{J}_{2N\times 2N}\bm{v}~=~\pm\bm{v}$. We can now choose half of the column vectors to be $\bm{v}_{i}~=~(\bm{a}_{i},\bm{a}^{f}_{i})^{T}$ where the superscript $f$ denotes that the sub-vector is flipped from left-to-right (i.e. $\bm{a}^{f}_{i}~=~\bm{J}_{N\times N}\bm{a}_{i}$). Since the second set of vectors are constructed via $\bm{\Omega}\bm{v}_{i}~=~(-\bm{a}^{f}_{i},\bm{a}_{i})^{T}$ we already have the set of antisymmetric vectors. We thus find that the symplectic orthogonal matrices for frequency symmetric pumps are
    \begin{align}
        \bm{O} = \begin{pmatrix}
            0_{N\times 1} & \bm{a}_{1} &  & 0_{N\times 1} & -\bm{a}^{f}_{1}  &  \\
            \bm{a}_{1} & 0_{N\times 1} &\ldots & \bm{a}^{f}_{1}  & 0_{N\times 1} &\ldots \\
            0_{N\times 1} & \bm{a}^{f}_{1} &  & 0_{N\times 1} & \bm{a}_{1} &  \\
            -\bm{a}^{f}_{1}  & 0_{N\times 1} &  & \bm{a}_{1} & 0_{N\times 1} &  \\
        \end{pmatrix},
    \end{align}
    \begin{align}
        \tilde{\bm{O}} = \begin{pmatrix}
            0_{N\times 1} & \bm{a}^{f}_{1}  &  & 0_{N\times 1} & -\bm{a}_{1} &  \\
            \bm{a}^{f}_{1}  & 0_{N\times 1} &\ldots & \bm{a}_{1} & 0_{N\times 1} &\ldots \\
            0_{N\times 1} & \bm{a}_{1} &  & 0_{N\times 1} & \bm{a}^{f}_{1}  &  \\
            -\bm{a}_{1} & 0_{N\times 1} &  & \bm{a}^{f}_{1}  & 0_{N\times 1} &  \\
        \end{pmatrix}.
    \end{align}    
    By comparing the first vectors of each, we see that both the real and imaginary parts are flips of each other. We also see that for one vector, the real and imaginary parts are also flips of each other. This is not quite the behaviour shown in Fig.~\ref{fig:modes}, however, recall that for plotting purposes we chose to be in a gauge where the imaginary part of the Schmidt mode is $0$ at the origin. Furthermore, the free-propagation phases are also removed in these plots(as outlined in reference~\cite{quesada2020theory}). Numerically, if one obtains the Schmidt modes above and applies the proper rotations to be in the same gauge, we obtain the results presented in Fig.~\ref{fig:modes}.
    
    Having obtained the Bloch-Messiah decomposition for an unpoled single pass structure and having looked at a more specific case, we now consider how poling possibly affects the results.
    
    %%%%%%%%%%%%%%%%%%%%%%%%%%%%%%%%%%%%%%%%%%%%%%%%%%%%%%%%%%%%%%%%%%%%%%%%
	%%%%%%%%%%%%%%%%%%%%%%%%%%%%%%%%%%%%%%%%%%%%%%%%%%%%%%%%%%%%%%%%%%%%%%%%
	\subsection{Single pass with quasi-phase-matching}\label{sec:qpm}
	%%%%%%%%%%%%%%%%%%%%%%%%%%%%%%%%%%%%%%%%%%%%%%%%%%%%%%%%%%%%%%%%%%%%%%%%
	%%%%%%%%%%%%%%%%%%%%%%%%%%%%%%%%%%%%%%%%%%%%%%%%%%%%%%%%%%%%%%%%%%%%%%%%

    We now consider the simplest type of poling which gives rise to QPM. This type of poling is represented in Fig.~\ref{fig:model}(b) where all the segments are of equal length ($z_{i}\equiv z$). To construct the full Heisenberg-picture propagator, we need to consider what happens when we choose the poling function $g(z)=-1$. In this case, we simply have that $\bm{F}~\rightarrow~-\bm{F}$ which in turn makes $\bm{A}~\leftrightarrow~-\bm{A}^{T}$. The propagator for a segment of length $\Delta z$ with negative poling is thus
    \begin{align}
        \bm{S}_{-} &= \bm{B}    \begin{pmatrix}
        e^{-\Delta z\bm{A}^{T}} & 0_{2N\times 2N}  \\
         0_{2N\times 2N} & e^{\Delta z\bm{A}}  \\
        \end{pmatrix} \bm{B}^{T},
    \end{align}
    where the $-$ subscript denotes that the poling function is negative. For the full propagator, we assume without loss of generality that the poling is symmetric (i.e. we pole with an odd number of domains) and we begin(and end) with positive poling. For a large number of domains, the solution should not depend on $\pm 1$ domains at the beginning or end of the waveguide. In this case, absorbing $\Delta z$ into $\bm{A}$, the propagator takes the form
    \begin{align}
        \bm{S} = \bm{B}\begin{pmatrix}
            e^{\bm{A}}e^{-\bm{A}^{T}}\cdots e^{-\bm{A}^{T}}e^{\bm{A}} & 0_{2N\times 2N} \\
            0_{2N\times 2N} & e^{-\bm{A}^{T}}e^{\bm{A}}\cdots e^{\bm{A}}e^{-\bm{A}^{T}}
        \end{pmatrix}   \bm{B}^{T}.
    \end{align}
    Clearly the two products are not symmetric, however, we can easily symmetrize the product by again multiplying each block by $\bm{X}$. To see how this works, recall that from the matrix structure of $e^{\bm{A}}$ and $e^{-\bm{A}^{T}}$ we have that the products $\bm{X}e^{\bm{A}}$ and $\bm{X}e^{\bm{-A^{T}}}$ are all symmetric. Therefore, we have that $\bm{X}e^{\bm{A}} = e^{\bm{A}^{T}}\bm{X}$ and $\bm{X}e^{\bm{-A^{T}}} =e^{-\bm{A}}\bm{X}$. We now consider
    \begin{align}
        \left( \bm{X}e^{\bm{A}}e^{-\bm{A}^{T}}\cdots e^{-\bm{A}^{T}}e^{\bm{A}}  \right)^{T} &= e^{\bm{A}^{T}}e^{-\bm{A}}\cdots e^{-\bm{A}}e^{\bm{A}^{T}}\bm{X} \nonumber\\
        &=e^{\bm{A}^{T}}e^{-\bm{A}}\cdots e^{-\bm{A}}\bm{X}e^{\bm{A}}\nonumber\\
        &=e^{\bm{A}^{T}}e^{-\bm{A}}\cdots \bm{X}e^{-\bm{A}^{T}}e^{\bm{A}}\nonumber\\
        &\cdots\nonumber\\
        &=\bm{X}e^{\bm{A}}e^{-\bm{A}^{T}}\cdots e^{-\bm{A}^{T}}e^{\bm{A}}
    \end{align}
    and therefore $\bm{X}e^{\bm{A}}e^{-\bm{A}^{T}}\cdots e^{-\bm{A}^{T}}e^{\bm{A}}$ is symmetric. All the previous arguments for the single unpoled case also apply here and so everything else follows for the Schmidt modes. We now wish to consider more arbitrary poling for more exotic phase-matching functions.
    %%%%%%%%%%%%%%%%%%%%%%%%%%%%%%%%%%%%%%%%%%%%%%%%%%%%%%%%%%%%%%%%%%%%%%%%
	%%%%%%%%%%%%%%%%%%%%%%%%%%%%%%%%%%%%%%%%%%%%%%%%%%%%%%%%%%%%%%%%%%%%%%%%
	\subsection{Single pass with arbitrary poling}
    \label{sec:arbpoling} 
	%%%%%%%%%%%%%%%%%%%%%%%%%%%%%%%%%%%%%%%%%%%%%%%%%%%%%%%%%%%%%%%%%%%%%%%%
	%%%%%%%%%%%%%%%%%%%%%%%%%%%%%%%%%%%%%%%%%%%%%%%%%%%%%%%%%%%%%%%%%%%%%%%%

    For more arbitrary poling functions, we cannot obtain analytic relations as specific as the no-poling or QPM poling, however, we can still obtain some information concerning the Schmidt modes. We assume without loss of generality that the poling starts with a positive segment and ends in a negative segment. Any of the arguments presented do not depend on the beginning and end structure. The total propagator can be considered as being a product of subsequent positively and negatively poled segments
    \begin{align}
        \bm{S} = \prod_{n} \bm{S}_{-}(z_{n+1})\bm{S}_{+}(z_{n}),
    \end{align}
    where $n$ goes from 1 to $N/2-1$ in step of 2, $N$ being the total number of segments. We can express the propagator as
    \begin{align}
        \bm{S} &= \bm{B}    \begin{pmatrix}
        \bm{\mathcal{A}} & 0_{2N\times 2N}  \\
         0_{2N\times 2N} & \bm{\mathcal{A}}^{-T}  \\
        \end{pmatrix} \bm{B}^{T},
    \end{align}
    where
    \begin{align}\label{eq:arbpolingA}
        \bm{\mathcal{A}} = e^{-z_{N/2}\bm{A}^{T}}e^{z_{N/2-1}\bm{A}}\cdots e^{-z_{2}\bm{A}^{T}}e^{z_{1}\bm{A}},
    \end{align}
    and $-T$ is understood to be the inverse-transpose. Now $\mathcal{\bm{A}}$ is  non-symmetric and without having knowledge of the poling we cannot symmetrize it as simply as in the no-poling case. However, from the singular-value decomposition, we can still obtain some information about the Schmidt modes. Since $\bm{\mathcal{A}}$ is real, the singular-value decomposition is given by
    \begin{align}
        \bm{\mathcal{A}} = \bm{\mathcal{M}}\bm{\mathcal{D}}\bm{\mathcal{V}}^{T},
    \end{align}
    where both $\bm{\mathcal{M}}$ and $\bm{\mathcal{V}}$ are real and orthogonal, and $\bm{\mathcal{D}}$ is diagonal with positive entries. From this definition, we find that
    \begin{align}
        \bm{\mathcal{A}}^{-T} = \bm{\mathcal{M}}\bm{\mathcal{D}}^{-1}\bm{\mathcal{V}}^{T}.
    \end{align}
    We can then identify the symplectic orthogonal matrices as
    \begin{align}
        \bm{O} &=\bm{B}\begin{pmatrix}
            \bm{\mathcal{M}} & 0_{2N\times 2N}  \\
            0_{2N\times 2N} & \bm{\mathcal{M}}  \\
        \end{pmatrix}\bm{W},\\
        \tilde{\bm{O}} &=\bm{B}\begin{pmatrix}
            \bm{\mathcal{V}} & 0_{2N\times 2N}  \\
            0_{2N\times 2N} & \bm{\mathcal{V}}  \\
        \end{pmatrix}\bm{W}.
    \end{align}
    The modes of $\bm{\mathcal{M}}$($\bm{\mathcal{V}}$) are related to the left(right) eigenvectors of $\bm{\mathcal{A}}\bm{\mathcal{A}}^{T}$. The case where $\bm{\mathcal{A}}=\bm{\mathcal{A}}^{T}$ implies $\bm{\mathcal{M}}=\bm{\mathcal{V}}$. Therefore, unless we have a very specific poling which gives rise to a symmetric $\bm{\mathcal{A}}$, the input and output Schmidt modes will be different. If we choose the pump to be frequency symmetric then since $\bm{\mathcal{A}}$ is comprised of products of centrosymmetric matrices, it too is centrosymmetric and therefore so is $\bm{\mathcal{A}}\bm{\mathcal{A}}^{T}$. This tells us that the modes of $\bm{\mathcal{M}}$ and $\bm{\mathcal{V}}$ will have the same flip structure as in the no-poling case, which gives rise to the behaviour shown in Fig.~\ref{fig:modes}.

    %%%%%%%%%%%%%%%%%%%%%%%%%%%%%%%%%%%%%%%%%%%%%%%%%%%%%%%%%%%%%%%%%%%%%%%%
	%%%%%%%%%%%%%%%%%%%%%%%%%%%%%%%%%%%%%%%%%%%%%%%%%%%%%%%%%%%%%%%%%%%%%%%%
	\subsection{Double pass with arbitrary poling and polarization swap}
    \label{sec:doublepass}
	%%%%%%%%%%%%%%%%%%%%%%%%%%%%%%%%%%%%%%%%%%%%%%%%%%%%%%%%%%%%%%%%%%%%%%%%
	%%%%%%%%%%%%%%%%%%%%%%%%%%%%%%%%%%%%%%%%%%%%%%%%%%%%%%%%%%%%%%%%%%%%%%%%

    We now consider the double pass structure portrayed in Fig.~\ref{fig:model}(c). Once the beams pass through the crystal, they are reflected back to pass a second time. The beams pass through a quarter-wave plate twice which swaps the polarization of the signal and idler beams. At the level of the equations of motion, we have that for the second pass $v_{S}\leftrightarrow v_{I}$, the signal and idler beams swap velocities. For the second pass, we then have that $\kappa\rightarrow -\kappa$ which makes $\bm{G}\rightarrow-\bm{G}$ and therefore $\bm{A}\leftrightarrow \bm{A}^{T}$. For the second pass, we also have that the domain configuration is flipped since the modes are passing through from the end back to the beginning. Using the same assumptions as in the previous section (sec.~\ref{sec:arbpoling}), we find that the propagator for the second pass is
    \begin{align}
        \bm{S}_{2} &= \bm{B}    \begin{pmatrix}
        \bm{\mathcal{A}}^{T} & 0_{2N\times 2N}  \\
         0_{2N\times 2N} & \bm{\mathcal{A}}^{-1}  \\
        \end{pmatrix} \bm{B}^{T},
    \end{align}
    where $\bm{\mathcal{A}}$ is given by Eq.~\ref{eq:arbpolingA} and the subscript $2$ denotes that it is the second pass. Using the same conventions for the singular-value decomposition, we have that the symplectic orthogonal matrices for the second pass are
    \begin{align}
        \bm{O}_{2} &=\bm{B}\begin{pmatrix}
            \bm{\mathcal{V}} & 0_{2N\times 2N}  \\
            0_{2N\times 2N} & \bm{\mathcal{V}}  \\
        \end{pmatrix}\bm{W},\\
        \tilde{\bm{O}}_{2} &=\bm{B}\begin{pmatrix}
            \bm{\mathcal{M}} & 0_{2N\times 2N}  \\
            0_{2N\times 2N} & \bm{\mathcal{M}}  \\
        \end{pmatrix}\bm{W}.
    \end{align}    
    From this we see that the input modes of the second pass are the output modes of the first pass and the output modes of the second pass are the input modes of the first. Combining both passes therefore gives a configuration where the input and output Schmidt modes are the same. The propagator for the full configuration is given by
    \begin{align}
        \bm{S} &= \bm{B}    \begin{pmatrix}
        \bm{\mathcal{A}}^{T}\bm{\mathcal{A}} & 0_{2N\times 2N}  \\
         0_{2N\times 2N} & \bm{\mathcal{A}}^{-1}\bm{\mathcal{A}}^{-T}  \\
        \end{pmatrix} \bm{B}^{T}\nonumber\\
         &= \bm{B}    \begin{pmatrix}
        \bm{\mathcal{V}}\bm{\mathcal{D}}^{2}\bm{\mathcal{V}}^{T} & 0_{2N\times 2N}  \\
         0_{2N\times 2N} &  \bm{\mathcal{V}}\bm{\mathcal{D}}^{-2}\bm{\mathcal{V}}^{T}  \\
        \end{pmatrix}\bm{B}^{T},
    \end{align}
    where the input and output Schmidt modes are given by
    \begin{align}
        \bm{O}=\tilde{\bm{O}} &=\bm{B}\begin{pmatrix}
            \bm{\mathcal{V}} & 0_{2N\times 2N}  \\
            0_{2N\times 2N} & \bm{\mathcal{V}}  \\
        \end{pmatrix}\bm{W}.
    \end{align}
    Again, if we assume the pump to be frequency symmetric the modes of $\bm{\mathcal{V}}$ will have the same flip structure as before.

    We see that the double pass structure of Fig.~\ref{fig:model}(c) leads to input and output Schmidt modes that are equivalent and thus represents a perfect inline squeezer. This conclusion only holds in the ideal case where both passes are identical. We now want to numerically study how robust the double pass structure is to variations in gain between the two passes. 

    %%%%%%%%%%%%%%%%%%%%%%%%%%%%%%%%%%%%%%%%%%%%%%%%%%%%%%%%%%%%%%%%%%%%%%%%
	%%%%%%%%%%%%%%%%%%%%%%%%%%%%%%%%%%%%%%%%%%%%%%%%%%%%%%%%%%%%%%%%%%%%%%%%
	\subsubsection{Robustness of double pass structure}\label{app:fidelity}
	%%%%%%%%%%%%%%%%%%%%%%%%%%%%%%%%%%%%%%%%%%%%%%%%%%%%%%%%%%%%%%%%%%%%%%%%
	%%%%%%%%%%%%%%%%%%%%%%%%%%%%%%%%%%%%%%%%%%%%%%%%%%%%%%%%%%%%%%%%%%%%%%%%

    In a typical experimental setting, several things could occur in between the passes which would effectively lower the gain level in the second pass. Such effects could be caused by losses or by some phase buildup between the modes which causes the phase-matching condition to no longer be fully met. One could also imagine a double pass structure where each pass is a unique nonlinear region which could have variations in their nonlinear strengths due to the fabrication process, allowing for a possibly higher or lower gain in the second pass.

\begin{figure}[t]
		\includegraphics[width=0.75\linewidth]{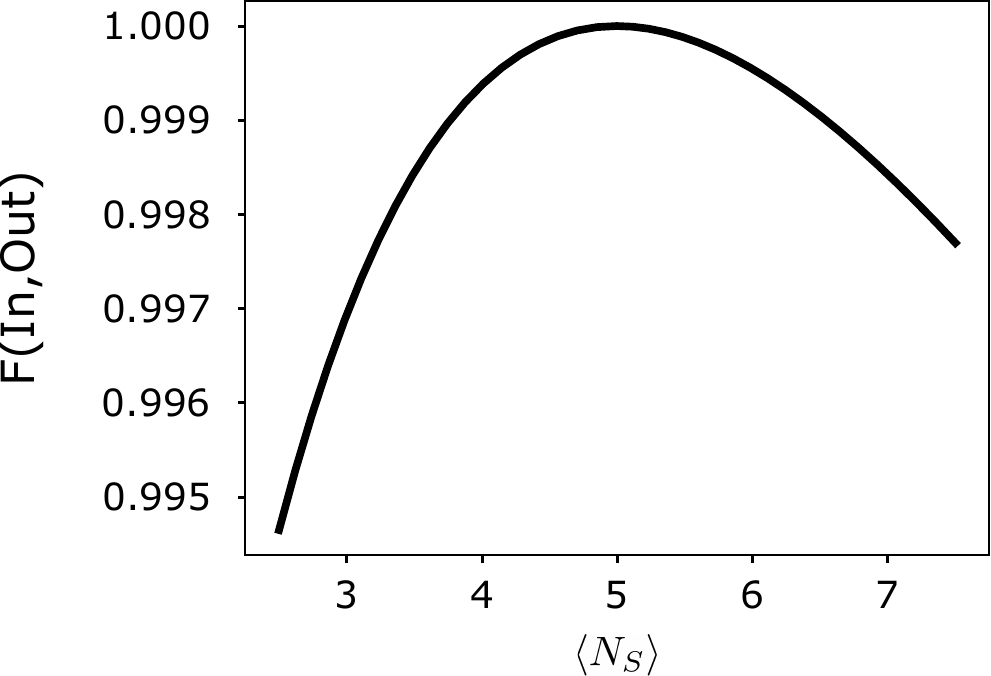}
		\caption{Fidelity between the input and output Schmidt modes for a double pass structure where the gain in the second pass is allowed to vary. We choose parameters such that when both passes have the same gain the average number of signal photons is $\langle  N_{S} \rangle = 5$. We allow for total variations of the average number of signal photons on the order of $\pm 2.5$. We see that the fidelity between input and output Schmidt modes remains very high.}
		\label{fig:fidelity}
	\end{figure}

    To study these effects, we consider a setup where we fix the level of gain in the first pass and allow the second pass to vary. We choose the parameters in such a way that when both passes give rise to the same gain, the average number of signal photons after the double pass is $\langle  N_{S} \rangle = 5$. We then vary the gain in the second pass such that the average number of signal photons after the double pass varies from $\langle  N_{S} \rangle = 2.5$ to $\langle  N_{S} \rangle = 7.5$ which is a difference in expected gain of $\pm 50\%$. We numerically compute the Bloch-Messiah decomposition as we sweep through the gain of the second pass and calculate the fidelity between the input Schmidt mode $\rho^{in}_{S}(\omega)$ and the output Schmidt mode $\rho^{out}_{S}(\omega)$ which we define as
    \begin{align}
        F(\text{In},\text{Out}) = \left|\int d\omega \rho^{out}_{S}(\omega) \left[ \rho^{in}_{S}(\omega) \right]^{*}\right|^{2}.
    \end{align}
    Figure~\ref{fig:fidelity} shows the fidelity as a function of the overall gain as characterized by $\langle N_{S }\rangle$. We see that even if the expected gain varies by $\pm 50\%$, the fidelity remains quite high ($>0.99$). This shows that the double pass structure is a robust perfect inline squeezer. All of the results above are for the SGVM regime, we now want to see if anything can be said for arbitrary signal and idler velocities.

   %%%%%%%%%%%%%%%
	%Figure 4
	%%%%%%%%%%%%%%%
    %%%%%%%%%%%%%%%
	%Figure 4
    %%%%%%%%%%%%%%%

    %%%%%%%%%%%%%%%%%%%%%%%%%%%%%%%%%%%%%%%%%%%%%%%%%%%%%%%%%%%%%%%%%%%%%%%%
	%%%%%%%%%%%%%%%%%%%%%%%%%%%%%%%%%%%%%%%%%%%%%%%%%%%%%%%%%%%%%%%%%%%%%%%%
	\subsection{Arbitrary signal and idler group velocity}\label{app:nosgvm}
	%%%%%%%%%%%%%%%%%%%%%%%%%%%%%%%%%%%%%%%%%%%%%%%%%%%%%%%%%%%%%%%%%%%%%%%%
	%%%%%%%%%%%%%%%%%%%%%%%%%%%%%%%%%%%%%%%%%%%%%%%%%%%%%%%%%%%%%%%%%%%%%%%%

    We consider how relaxing the assumption of SGVM modifies the approach to obtaining the Bloch-Messiah decomposition. In this case, we do not have that $\bm{G}(z)~=~-\bm{H}(z)$ and so we need to keep them distinct. As this adds complexity, we assume that the pump envelope function, $\beta_{P}(\omega)$, is frequency symmetric. Recall that this makes $\bm{F}(z)$ centrosymmetric. We also consider that we take both our frequency grids such that $\omega_{n}-\bar{\omega}_{S}$ and $\omega_{n}-\bar{\omega}_{I}$ take on values between $\pm \delta$ for some chosen $\delta$. This gives us that both $\bm{G}(z)$ and $\bm{H}(z)$ are anti-centrosymmetric and that all block-matrices are symmetric. As in the matched regime, we focus on a short segment of length $\Delta z$ where the poling function $g(z)=1$. The equations of motion are governed by the $z$-independent matrix
    \begin{align}
        \tilde{\bm{Q}}=\begin{pmatrix}
			0_{N\times N} & 0_{N\times N} & -\bm{G} & \bm{F} \\
			0_{N\times N} & 0_{N\times N} & \bm{F} & -\bm{H} \\
            \bm{G} & \bm{F} & 0_{N\times N} & 0_{N\times N}  \\
            \bm{F} & \bm{H} & 0_{N\times N} & 0_{N\times N}  \\
		\end{pmatrix},        
    \end{align}
    with matrix elements described above (Eqs.~\ref{eq:Gmat}-\ref{eq:Fmat}). The orthogonal matrix, $\bm{B}$ (Eq.~\ref{eq:Bmat}), does not transform the matrix $\tilde{\bm{Q}}$ into block-diagonal form. One needs to consider a different transformation of the form
    \begin{align}
        \bm{B}_{2}=\frac{1}{\sqrt{2}}\begin{pmatrix}
            0_{N\times N} & \mathbb{1}_{N\times N} & 0_{N\times N} & \bm{J}_{N\times N} \\
            \mathbb{1}_{N\times N} & 0_{N\times N} & \bm{J}_{N\times N} & 0_{N\times N} \\
            0_{N\times N} & -\bm{J}_{N\times N} & 0_{N\times N} & \mathbb{1}_{N\times N} \\
            -\bm{J}_{N\times N} & 0_{N\times N} & \mathbb{1}_{N\times N} &  0_{N\times N} \\
        \end{pmatrix},        
    \end{align}
    where $\mathbb{1}_{N\times N}$ is the identity matrix and $\bm{J}_{N\times N}$ is the exchange matrix. Under this transformation, we have that
    \begin{align}
        \bm{B}^{T}_{2}\tilde{\bm{Q}}\bm{B}_{2} = \begin{pmatrix}
            \bm{C} & 0_{2N\times 2N} \\
            0_{2N\times 2N} & -\bm{C}^{T}\\
        \end{pmatrix},
    \end{align}
    where
    \begin{align}
        \bm{C} = \begin{pmatrix}
            \bm{H}\bm{J}-\bm{J}\bm{H} & -\bm{J}\bm{F}-\bm{F}\bm{J} \\
            -\bm{J}\bm{F}-\bm{F}\bm{J} & \bm{G}\bm{J}-\bm{J}\bm{G} \\
        \end{pmatrix} \equiv \begin{pmatrix}
            \tilde{\bm{H}} & \tilde{\bm{F}} \\
            \tilde{\bm{F}} & \tilde{\bm{G}} \\
        \end{pmatrix}.
    \end{align}
    Note that if the pump is not frequency symmetric, the off-diagonal blocks to not vanish. Since the matrices $\bm{F}$, $\bm{G}$, and $\bm{H}$ are all symmetric, the blocks behave such that $\tilde{\bm{F}}^{T}=\tilde{\bm{F}}$, $\tilde{\bm{H}}^{T}=-\tilde{\bm{H}}$, and $\tilde{\bm{G}}^{T}=-\tilde{\bm{G}}$. Clearly $\bm{C}$ is not symmetric. We use a similar strategy as before, but now we consider a variation of the exchange matrix
    \begin{align}
        \tilde{\bm{J}} = \begin{pmatrix}
            \bm{J}_{N\times N} & 0_{N\times N} \\
            0_{N\times N} & \bm{J}_{N\times N}\\
        \end{pmatrix}.
    \end{align}
    From the structure of $\bm{C}$ we find that
    \begin{align}
        \tilde{\bm{J}}\bm{C}\tilde{\bm{J}}=\bm{C}^{T}.
    \end{align}
    As in the previous cases, we can then show that $\tilde{\bm{J}}e^{\Delta z \bm{C}}$ is symmetric. This allows us again to express it via its eigendecomposition
    \begin{align}
        \tilde{\bm{J}}e^{\Delta z \bm{C}} = \bm{\Gamma}\bm{\Lambda}\bm{\Gamma}^{T}
    \end{align}
    with $\bm{\Gamma}$ real and orthogonal and $\bm{\Lambda}$ real and diagonal. By taking the inverse-transpose we find that
    \begin{align}
        \tilde{\bm{J}}e^{-\Delta z \bm{C}^{T}} = \bm{\Gamma}\bm{\Lambda}^{-1}\bm{\Gamma}^{T}.
    \end{align}    
    Similarly to the SGVM regime, the Heisenberg-picture propagator can then be written as
    \begin{align}
        \bm{S}&=\bm{B}_{2} \begin{pmatrix}
            \tilde{\bm{J}}\bm{\Gamma} & \bm{0}  \\
             \bm{0} & \tilde{\bm{J}}\bm{\Gamma}  \\
        \end{pmatrix}\begin{pmatrix}
            \bm{\Lambda} & \bm{0}  \\
             \bm{0} & \bm{\Lambda}^{-1}  \\
        \end{pmatrix} \begin{pmatrix}
            \bm{\Gamma}^{T} & \bm{0}  \\
             \bm{0} & \bm{\Gamma}^{T}  \\
        \end{pmatrix}\bm{B}_{2}^{T}, 
    \end{align}
    where $\bm{0}=0_{2N\times 2N}$. We now need to show that we also have the proper degeneracy expected in the Bloch-Messiah decomposition. To do so, we consider an extended Pauli-Z matrix
    \begin{align}
        \bm{Z} = \begin{pmatrix}
            \mathbb{1}_{N \times N} & 0_{N \times N} \\
            0_{N \times N}  & -\mathbb{1}_{N \times N}\\
        \end{pmatrix}.
    \end{align}    
    From the structure of $\bm{C}$ we have that
    \begin{align}
        \bm{Z}\bm{C}\bm{Z} = -\bm{C}^{T}.
    \end{align}
    Now suppose $\bm{v}$ is an eigenvector of $\tilde{\bm{J}}e^{\Delta z \bm{C}}$ with eigenvalue $\lambda$ then we also know that it is an eigenvector of $\tilde{\bm{J}}e^{-\Delta z \bm{C}^{T}}$ with eigenvalue $\lambda^{-1}$. Now consider the vector $\bm{Z}\bm{v}$, we have that
    \begin{align}
        \tilde{\bm{J}}e^{\Delta z \bm{C}} \bm{Z}\bm{v} &=\tilde{\bm{J}}\bm{Z}\bm{Z}e^{\Delta z \bm{C}} \bm{Z}\bm{v}\nonumber\\
        &=\tilde{\bm{J}}\bm{Z}e^{-\Delta z \bm{C}^{T}}\bm{v}\nonumber\\
        &=\bm{Z}\tilde{\bm{J}}e^{-\Delta z \bm{C}^{T}}\bm{v}\nonumber\\
        &=\bm{Z}\lambda^{-1}\bm{v}\nonumber\\
        &=\lambda^{-1}\bm{Z}\bm{v}
    \end{align}
    and so $\bm{Z}\bm{v}$ is an eigenvector of $\tilde{\bm{J}}e^{\Delta z \bm{C}}$ with eigenvalue $\lambda^{-1}$. We have the required degeneracy but instead of appearing with a negative value as in the SGVM case, the diagonal matrix contains each value and its inverse. Since $\bm{Z}\bm{v}$ is also an eigenvalue $\tilde{\bm{J}}e^{-\Delta z \bm{C}^{T}}$ with eigenvalue $\lambda$ we can simply reorder the eigenvalues by placing all the inverses in the bottom diagonal block without having to reorder the eigenvectors. 
    
    The matrix $\bm{C}$ is not centrosymmetric so the eigenvectors will not have any flip structure as in the matched case. The arguments concerning QPM and arbitrary poling also hold here.  

    Interestingly, a double pass structure without SGVM will not have the same input and output Schmidt modes. Unlike the main text, the polarization swap takes $\bm{C}\rightarrow \bm{X}\bm{C}\bm{X}\neq \bm{C}^{T}$ and so the arguments do not hold. In fact, we can only say as much as the single pass with arbitrary poling case.
    
    We thus gather that to obtain a perfect inline squeezer, one needs to be working in the SGVM regime.

	%%%%%%%%%%%%%%%%%%%%%%%%%%%%%%%%%%%%%%%%%%%%%%%%%%%%%%%%%%%%%%%%%%%%%%%%
	%%%%%%%%%%%%%%%%%%%%%%%%%%%%%%%%%%%%%%%%%%%%%%%%%%%%%%%%%%%%%%%%%%%%%%%%
	\section{Conclusion}
	\label{sec:conc}
	%%%%%%%%%%%%%%%%%%%%%%%%%%%%%%%%%%%%%%%%%%%%%%%%%%%%%%%%%%%%%%%%%%%%%%%%
	%%%%%%%%%%%%%%%%%%%%%%%%%%%%%%%%%%%%%%%%%%%%%%%%%%%%%%%%%%%%%%%%%%%%%%%%

    In this paper, we have developed a method to obtain analytical properties for the Bloch-Messiah decomposition of the Heisenberg-picture propagators describing several different parametric waveguided squeezers. This allows us to be able to directly compare the input and output temporal modes to properly characterize the squeezed state generated when seeding the squeezer with non-vacuum states.

    In the SGVM regime, we find that single pass structures with and without poling lead to distinct input and output temporal modes. Furthermore, we find that these temporal modes have their real and imaginary parts swapped. By considering a frequency symmetric pump, we also find that the real and imaginary parts are flips of each other, which in the proper gauge explains the numerical results of Fig.~\ref{fig:modes}. Experimentally, one could imagine using a phase-shifter to correct the phases before sending the signal back into the single pass structure. However, for the pump pulse time-scales considered (femtoseconds) no such device exists. State of the art phase-shifters work up to the picoseconds regime. Furthermore, the single pass Schmidt mode structure is highly gain dependent\cite{houde2023sources} and it would require extensive calibration to ensure that the phase-shifter is properly correcting for the phase mismatch. Even if one could use a phase-shifter it would highly convolute the experimental design due to this calibration. Phase-shifters also introduce significant noise in the system which would severely affect the squeezing properties of the setup.  As such, the single pass configuration does not serve as a perfect inline squeezer.      

    On the other hand, we find that the double pass configuration leads to equivalent input and output modes. This is due to the polarization swap which causes the input(output) modes of the second pass to be the output(input) modes of the first pass. Combining both in turn leads to an overall squeezer with the same input and output modes. Hence, not only is the double pass configuration practical for quantum computation~\cite{houde2023sources}, it also serves as a perfect inline squeezer. In section~\ref{app:fidelity}, we numerically study how relaxing the assumption that both passes have the same level of gain affects the input and output temporal modes and find that the double pass structure is a robust perfect inline squeezer.  

    By relaxing the assumptions on the group velocities of the beams, we find that for the arbitrary case there is no configuration which leads to a perfect inline squeezer. In this case, the input and output modes are always distinct, unless the poling is possibly chosen in a very specific way to make the interaction matrix symmetric. Although it could be possible to achieve such poling, chances are that other properties such as spectral purity would be negatively affected. 

    The results presented are relevant to any setup which uses these kinds of squeezers~\cite{zhong2020exp1,zhong2021exp2,Reddy2017conversion, Reddy2017ramsey, Furasawa2014dynamicsqueezing, winnel2023deterministic, crescimanna2023seeding, roeder2023measurement}. Our work represents one of the first studies concerning how to obtain perfect inline squeezers in twin-beam systems as well as obtaining analytic relations between the input and output Schmidt modes of complex systems without the use of numerics. 
    
    Having developed this method, an interesting next step would be to try applying a similar formalism to other types of squeezers (i.e. type-\RomanNumeralCaps{1} SPDC) as well trying to extend the formalism to also include more complicated interactions such as self- and cross-phase modulation as well as loss.
	
	\section*{Acknowledgements}
	The authors acknowledge support from the Minist\`{e}re de l'\'{E}conomie et de l’Innovation du Qu\'{e}bec and the Natural Sciences and Engineering Research Council of Canada. This work has been funded by the European Union's Horizon Europe Research and Innovation Programme under agreement 101070700 project MIRAQLS. The authors thank W. McCutcheon for insightful discussions concerning Bloch-Messiah decomposition and its implementation in Python, J.E. Sipe and Guillaume Thekkadath for insightful discussions, and Karthik Chinni for providing feedback on the manuscript.

% Create the reference section using BibTeX:
\bibliography{main}

\end{document}